\documentclass[12pt,epsf]{article}

\newlength{\pubnumber} 

\catcode`\@=11
\@addtoreset{equation}{section}

\def\section{\@startsection{section}{1}{\z@}{3.5ex plus 1ex minus .2ex}
 {2.3ex plus .2ex}{\large\bf}}
\def\subsection{\@startsection{subsection}{2}{\z@}{2.3ex plus .2ex}
 {2.3ex plus .2ex}{\bf}}

\usepackage{graphicx}% Include figure files
\usepackage{dcolumn}% Align table columns on decimal point
\usepackage{bm}% bold math
\usepackage{supertabular}

\usepackage[margin=1in]{geometry}

\usepackage{amsmath}
\usepackage{amsfonts}
\usepackage{amssymb}
\usepackage{hyperref}
\def\beq{\begin{equation}}
\def\eeq{\end{equation}}
\def\beqn{\begin{eqnarray}}
\def\eeqn{\end{eqnarray}}

\def\ol{\overline}

\def\oG{\overline{G}}
\def\op{\overline{\phi}}
\def\os{\overline{\psi}}
\def\bs{\overline{\psi}}
\def\p{\phi}
\def\s{\psi}

\def\bb{{\mbf b}}
\def\b{{\bf b}}

\def\mbf{\mathbf}
\def\bone{{\mathbf 1}}

\def\b{{\bf b}}

\def\bone{{\mathbf 1}}

\def\bS{{\mathbf S}}
\def\mS{{\mathbf S}}
\def\bS{{\mathbf S}}
\def\bb{{\mathbf b}}
\def\mb{{\mathbf b}}

\def\bV{{\mathbf V}}

\def\inbar{\,\vrule height1.5ex width.4pt depth0pt}
\def\IT{\relax\hbox{$\inbar\kern-.3em{\rm T}$}}
\def\IS{\relax\hbox{$\inbar\kern-.3em{\rm S}$}}
\def\IC{\relax\hbox{$\inbar\kern-.3em{\rm C}$}}
\def\IQ{\relax\hbox{$\inbar\kern-.3em{\rm Q}$}}
\def\IR{\relax{\rm I\kern-.18em R}}
 \font\cmss=cmss10 \font\cmsss=cmss10 at 7pt

\def\IZ{\relax\ifmmode\mathchoice
 {\hbox{\cmss Z\kern-.4em Z}}{\hbox{\cmss Z\kern-.4em Z}}
 {\lower.9pt\hbox{\cmsss Z\kern-.4em Z}}
 {\lower1.2pt\hbox{\cmsss Z\kern-.4em Z}}\else{\cmss Z\kern-.4em Z}\fi}

\def\Io{\relax\ifmmode\mathchoice
 {\hbox{\cmss 1\kern-.4em 1}}{\hbox{\cmss 1\kern-.4em 1}}
 {\lower.9pt\hbox{\cmsss 1\kern-.4em 1}}
{\lower1.2pt\hbox{\cmsss 1\kern-.4em 1}}\else{\cmss 1\kern-.4em 1}\fi}

\begin{document}

\begin{titlepage}
\setcounter{page}{1}
%%%%%%%%% PREPRINT NUMBERS %%%%%%%%%%%%%%%
\rightline{BU-HEPP-08-19}
\rightline{CASPER-08-08}
\rightline{\tt }

\vspace{.06in}
\begin{center}
{\Large \bf On a NAHE Variation}
\vspace{.12in}

{\large Jared Greenwald$^1$\footnote{jared\_greenwald@baylor.edu},
Douglas Moore$^1$\footnote{douglas\_moore1@baylor.edu},
        Kristen Pechan$^{1,2}$\footnote{astarforpele@gmail.com},
        Timothy Renner$^1$\footnote{timothy\_renner@baylor.edu},\\ 
        Tibra Ali$^{1,3}$\footnote{tali@perimeterinstitute.ca}, and
        Gerald Cleaver$^1$\footnote{gerald\_cleaver@baylor.edu}}
\\
\vspace{.12in}
{\it $^1$Center for Astrophysics, Space Physics \& Engineering Research\\
     Department of Physics, One Bear Place \#97316, Baylor University\\
     Waco, TX 76798-7316\\
\smallskip
$^2$ Mitchell Institute, \\
Texas A\&M University\\
College Station, TX 77843-4242\\
\smallskip
$^3$Perimeter Institute for Theoretical Physics\\
31 Caroline Street North,\\
Waterloo, Ontario\\
Canada N2L 2Y5\\
     }
\vspace{.06in}
\end{center}

\begin{abstract}

We present a variation of the NAHE-basis for free fermionic heterotic string models. By rotating some of the boundary conditions of the NAHE periodic/anti-periodic fermions \{$y^m$, $\overline{y}^m$, $w^m$, $\overline{w}^m$,\}, for $m=$ 1 to 6,
associated with the six compact dimensions of a bosonic lattice/orbifold model, we show an additional method for enhancing the standard NAHE gauge group of $SO(10)$ back to $E_6$. 
This rotation transforms $(SO(10)\otimes SO(6)^3)_{\rm obs} \otimes (E_8)_{\rm hid}$ into $(E_6 \otimes U(1)^5)_{\rm obs} \otimes SO(22)_{\rm hid}$.
When $SO(10)$ is enhanced to $E_6$ in this manner, 
the $i^{\rm th}$ MSSM matter generation in the $SO(10)$ $\mathbf{16}_i$ rep, originating in the twisted basis vector $\mathbf{b}_i$, recombines with both its associated untwisted MSSM Higgs in a $\mathbf{10}_i$ rep and an untwisted non-Abelian singlet $\phi_i$, to form a $\mathbf{27}_i$ rep of $E_6$. 
Beginning instead with the $E_6$ model, the inverse transformation of the fermion boundary conditions corresponds to partial GUT breaking via boundary rotation. 
%Thus, comparison of the standard NAHE basis with our variation demonstrates %how twisting of the six compact spacetime dimensions can break $E_6$ to the %$SO(10)\otimes U(1)$ of NAHE. 

Correspondence between free fermionic models with $\IZ_2\otimes \IZ_2$ twist (especially of the NAHE class) and orbifold models with a similar twist has received further attention recently. Our NAHE variation also involves a $\IZ_2\otimes \IZ_2$ twist and offers additional understanding regarding the free fermion/orbifold correspondence. Further, models based on this NAHE variation offer some different phenomenological features compared to NAHE-based models. In particular, the more compact 
$\IZ_2\otimes \IZ_2$ twist of the NAHE variation offers a range of mirror models not possible from NAHE-based models. Examples of such models are presented.
\end{abstract}
\end{titlepage}
\setcounter{footnote}{0}

%*************************************************************************************
\section{NAHE Variation with a Geometric Twist}\label{sec:ffm1}

The parameter space of the weakly coupled free fermionic heterotic string (WCFFHS) \cite{fff1,fff2} region of the string/M landscape has proven to be rich in quasi-realistic models containing the Minimal Supersymmetric Standard Model or its extensions.  
The WCFFHS region has produced a vast range of quasi-realistic (Near-)MSSM-like models \cite{nmssm,freegen,mshsm,optun}, semi-GUT models \cite{sguts,cnfv,psm2,acfkr}, and GUT models \cite{fsu5}. 
The majority of these models are constructed as extensions of the NAHE (Nanopoulos, Antoniadis, Hagelin, Ellis) set \cite{nahe}, with the $5$ basis vectors of the NAHE set as their common core. Within the five basis vectors of the NAHE set, the 12 real free fermions representing the 6 compactified bosonic directions have boundary condition vectors equivalent to a $T^6/\IZ_2\otimes \IZ_2$ orbifold twist. While basis vector extensions to the NAHE set may or may not break this $\IZ_2\otimes \IZ_2$ symmetry, the quasi-realistic models consistently do not.

The phenomenological fruitfulness of the WCFFHS region of the string landscape continues to inspire attention. Recent random searches  of the region have been performed \cite{af1,kd,acfkr} and systematic searches are underway \cite{rch,rch2,rch3,rch4}. 
Distribution functions of various phenomenological features have been computed and are being further refined by the systematic searches. 

Also of current focus is the correspondence between free fermionic and orbifold models \cite{rd1,ss1,ss2,ss3,ft}. In \cite{rd1} a complete classification was obtained for orbifolds of the form $X/G$, with $X$ the product of three elliptic curves and $G$ an Abelian extension of a group of $\IZ_2\otimes \IZ_2$ twists acting on $X$. This includes $T^6/\IZ_2\otimes \IZ_2$ orbifolds.
Each such orbifold was shown to correspond to a free fermionic model with geometric interpretation. 
The NAHE basis and certain model extensions were shown to have geometric interpretation and thus, have orbifold equivalences. However, the general class of quasi-realistic models with a NAHE basis were shown not to have geometric interpretation--specifically, their Hodge numbers were not reproducible by any orbifold $X/G$. In other words, the beyond-NAHE basis vectors necessary to yield a quasi-realistic model (by reducing the number of copies of each generation from 16 to 1 and breaking $SO(10)$ to a viable sub-group\footnote{$SO(10)$ must be broken via Wilson loop effects of basis vectors rather than by GUT Higgs, since adjoint or higher dimension scalars are not possible in Ka\v c-Moody rank one models.})
consistently break the $T^6/\IZ_2\otimes \IZ_2$ symmetry in a manner that also eliminates geometric interpretation.

The non-geometric feature of the quasi-realistic WCFFHS models inspired us to investigate variations of the NAHE set that might allow for quasi-realistic models with geometric interpretation, particularly with geometric $T^6/\IZ_2\otimes \IZ_2$ interpretation. In the next section, we construct a NAHE variation of this form by rotating (interchanging) the boundary conditions of a subset of the 12 real fermions in two of the twisted sectors. We then explore some of the phenomenological aspects of our new model class, especially in comparison to those of the NAHE class.

\section{Construction and Phenomenology of the NAHE Variation}
 
The NAHE basis set contains 5 basis vectors: The all-periodic sector $\bone$ (present in all free fermionic models), the supersymmetry generating sector $\bS$, and the three generation sectors $\bb_{i=1,2,3}$:
The NAHE set is depicted in Table 1 below 
(in which a ``1'' denotes a periodic fermion and a ``0'' denotes an antiperiodic fermion),
which highlights its cyclic permutation symmetry. In Table 1, 
the $(y,\, w)^m$, for $m=1$ to $6$, 
are the six pairs of real fermions that replace the left-moving bosonic scalar fields 
$X_m$ for the six compactified directions and the corresponding 
$(\overline{y},\, \overline{w})^m$ are the six pairs of real fermions that replace the left-moving 
$\overline{X}_m$. All other fermions in Table 1 are complex.

The gauge group resulting from the NAHE set is
$SO(10)\times SO(6)^3\times E_8$ with $N=1$ space--time supersymmetry.
The matter content is 48 spinorial $\mbf {16}$'s of $SO(10)$ matter states, coming from sixteen copies from each sector $\mb_1$, $\mb_2$ and $\mb_3$. The sixteen copies in each sector are composed of
2 copies of $(\mbf{16},\mbf{4}_i)$ reps and 2 copies of $(\overline{\mbf{16}},\overline{\mbf 4}_i)$
reps of $SO(10)\times SO(6)_i$ (i=1,...,3 for each of the three $SO(6)$'s). The untwisted sector also contains six copies of a pair of Higgs for each generation 
in the form of $(\mbf{10},\mbf{6}_i)$ reps of $SO(10)\times SO(6)_i$, in addition to a single $(\mbf{6}_i,\mbf{6}_j)$
rep of $SO(6)_i\otimes SO(6)_j$, for each case of $i,j\in \{1,2,3\}$ and $i\neq j$.
In a real basis of the $\overline{y}$ and $\overline{w}$, the generators 
of $SO(6)_1$ are $(\overline{\eta}^1, \overline{y}^1,\, \overline{y}^2,\, \overline{\\w}^5, \overline{w}^6)$;
of $SO(6)_2$ are $(\overline{\eta}^2, \overline{y}^3,\, \overline{y}^4,\, \overline{y}^5, \overline{y}^6)$; and 
of $SO(6)_3$ are $(\overline{\eta}^2, \overline{w}^1,\, \overline{w}^2,\, \overline{w}^3, \overline{w}^4)$.

The three sectors $\mb_1$, $\mb_2$ and $\mb_3$ are
the three unique twisted sectors of the corresponding $\IZ^a_2\times \IZ^b_2$
orbifold compactification. The $\IZ^a_2\times \IZ^b_2$ acts on the 
$(y,\, w)_i$ and $(\overline{y},\, \overline{w})_i$ in the $\bb_i$ according to
\beqn
\IZ^a_2:&&\ (y,\, \overline{y})^{m=3\, ... , 6} \rightarrow \ (y+1,\, \overline{y}+1)^m \; 
({\rm mod}\, 2)\label{za1}\\
\IZ^b_2:&&\ (y,\, \overline{y})^{m=1,\, 2};\,  
(w,\, \overline{w})^{n=5,\, 6}   \rightarrow \ (y+1,\, \overline{y}+1)^{m};
\, (w+1,\, \overline{w}+1)^{n}\,  ({\rm mod}\, 2).\label{za2}     
\eeqn
Thus, $\b_1$ is a $\IZ_2^a$ twisted sector; $\b_2$ is a $\IZ_2^b$ twisted sector, 
and   $\b_3 + \bone$ is a $\IZ_2^a\otimes \IZ_2^a$ twisted sector.
The $\IZ_2^a\times \IZ_2^b$ NAHE orbifold is special
precisely because of the existence of three twisted sectors (one per generation) that have a permutation symmetry with respect to the horizontal $SO(6)^3$
symmetries. This symmetry enables $\b_1 + \b_2 + \b_3 + \bone$ to generate the massless sector that produces the spinor components of the hidden sector $E_8$ gauge group.

As discussed previously, the NAHE set is common to a large class of three generation
free fermionic models. Model construction proceeds by adding to the
NAHE set three or four additional boundary condition basis vectors
which simultaneously break $SO(10)$ to one of its subgroups, $SU(5)\times U(1)$,
$SO(6)\times SO(4)$ or $SU(3)\times SU(2)\times U(1)^2$,
and reduce the number of generations to
three chiral, one from each of the sectors $\mb_1$, $\mb_2$ and $\mb_3$.
The various three generation models differ in their
detailed phenomenological properties based on the specific assignment of boundary condition
basis vector for the internal world--sheet fermions
$\{y,w\vert{\overline y},{\overline{w}}\}^{1,\cdots,6}$.
This is one reason for our interest in examining the properties of a new class of models based on a 
NAHE variation for which some of the boundary conditions of the
$\{y,w\vert{\overline y},{\overline{w}}\}^{1,\cdots,6}$  
are exchanged.

\vspace{0.15cm}
~~~~~{{\bf Table 1. The NAHE Set}}
{\large
{
\beqn
 &&\begin{tabular}{||c|c||c|c|c|c|c|c|c|c|c||}
 \hline \hline
 \textbf{Sec}&\textbf{N} & $\psi^\mu$ & ${x^{12}}$ & ${x^{34}}$ & ${x^{56}}$ &
        $\overline{\psi}^{1,...,5} $ &
        {$\overline{\eta}^1$}&
        {$\overline{\eta}^2$}&
        {$\overline{\eta}^3$}&
        $\overline{\phi}^{1,...,8} $ \\
\hline
\hline
      $\bone$&2 &  1 & 1&1&1 & 1,...,1 & 1 & 1 & 1 & 1,...,1 \\
      \hline
        $\mS$&2 &  1 & {1}&{1}&{1} & 0,...,0 & 0 & 0 & 0 & 0,...,0 \\
\hline
  {${\mb}_1$} &2&  1 & {1}&0&0 & 1,...,1 & {1} & 0 & 0 & 0,...,0 \\
  \hline
  {${\mb}_2$} &2&  1 & 0&{1}&0 & 1,...,1 & 0 & {1} & 0 & 0,...,0 \\
  \hline
  {${\mb}_3$} &2&  1 & 0&0&{1} & 1,...,1 & 0 & 0 & {1} & 0,...,0 \\
  \hline \hline
\end{tabular}
   \nonumber\\
   ~  &&  ~ \nonumber\\
     &&\begin{tabular}{||c|c||c|c|c|c|c|c|}
     \hline \hline
 \textbf{Sec}&\textbf{N}&     {$y^{3,...,6}$}  &
        {${\overline y}^{3,...,6}$}  &
        {$y^{1,2},w^{5,6}$}  &
        {${\overline y}^{1,2},\overline{w}^{5,6}$}  &
        {$w^{1,...,4}$}  &
        {$\overline{w}^{1,...,4}$}   \\
\hline
\hline
   $\bone$ &2 & 1,...,1 & 1,...,1 & 1,...,1 & 1,...,1 & 1,...,1 & 1,...,1 \\
   $\mS$&2    & 0,...,0 & 0,...,0 & 0,...,0 & 0,...,0 & 0,...,0 & 0,...,0 \\
\hline
{${\mb}_1$}&2 & {1,...,1} & {1,...,1} & 0,...,0 & 0,...,0 & 
                                          0,...,0 & 0,...,0 \\
                                          \hline
{${\mb}_2$}&2 & 0,...,0 & 0,...,0 & {1,...,1} & {1,...,1} & 
                                          0,...,0 & 0,...,0 \\
                                          \hline
{${\mb}_3$} &2& 0,...,0 & 0,...,0 & 0,...,0 & 0,...,0 & 
                                     {1,...,1} & {1,...,1} \\
                                     \hline \hline
\end{tabular}
\nonumber
\eeqn
}}

The NAHE variation under discussion is produced by exchanging some of the periodic and anti-periodic boundary conditions in the second and third generation sectors, as shown in Table 2: 
In $\bb_2$ the boundary conditions of $(y,\overline{y})^{m=5,6}$ and $(w,\overline{w})^{m=5,6}$ are interchanged 
and in $\bb_3$ 
the boundary conditions of $(y,\overline{y})^{m=1,2,3,4}$ and $(w,\overline{w})^{m=1,2,3,4}$ are interchanged. 
Under this exchange, both $\IZ^a$ and $\IZ^b$ now induce twists solely among the $(y,\, \overline{y})^m$ and no longer among the 
$(w,\, \overline{w})^m$. Further, $\IZ^a \otimes \IZ^b$ now corresponds exactly to $\bb_3$, rather than to 
$\bb_3 + \bone$. The effect of the exchanged boundary conditions for the $\IZ^a$ and $\IZ^b$ twists is very non-trivial. 

\vspace{0.15cm}
~~~~~{{\bf Table 2. A Variation on the NAHE Set}}
{\large
{
\beqn
    &&\begin{tabular}{||c|c||c|c|c|c|c|c|c|c||}
    \hline \hline
 \textbf{Sec}&\textbf{N}&     {$y^{1,2}$}         &
        {${\overline y}^{1,2}$}  &
        {$y^{3,4}$}         &
        {${\overline y}^{3,4}$}  &
        {$y^{5,6}$}         &
        {${\overline y}^{5,6}$}  &
        {$w^{1,...,6}$}&
        {$\overline{w}^{1,...,6}$}   \\
\hline
\hline
{${\mb}_1$}&2 & 0,0 & 0,0 & 1,1 & 1,1 & 1,1 & 1,1 &
                                          0,...,0 & 0,...,0 \\
                                          \hline
{${\mb}_2$}&2 & 1,1 & 1,1 & 0,0 & 0,0 & 1,1 & 1,1 &
                                          0,...,0 & 0,...,0 \\
                                          \hline
{${\mb}_3$}&2 & 1,1 & 1,1 & 1,1 & 1,1 & 0,0 & 0,0 &
                                          0,...,0 & 0,...,0 \\
                                          \hline \hline
\end{tabular}
\nonumber
\eeqn
}}

The observable gauge group is enhanced to $E_6 \otimes U(1)^5$ and the hidden sector gauge group transforms into $SO(22)$. 
The change in gauge group occurs because now it is the combination of $\bS + \bb_1 + \bb_2 + \bb_3$,  rather than of $\bone + \bb_1 + \bb_2 + \bb_3$, that forms a massless spinor gauge group sector. Thus, in the NAHE variation there is a massless spinor sector involving the five complex $\overline{\psi}$ and the three complex $\overline{\eta}$ observable sector fermions rather than the eight complex $\overline{\phi}$ hidden sector fermions. This massless spinor sector enhances the $SO(10)$ symmetry into $E_6$. The enhancement is into $E_6$ rather than $E_8$ because of the GSO constraints the $\bb_i$ basis vectors place on the $\overline{\eta}^i$ spinors. 

The trace component of the 3 complex $\overline{\eta}$ fermions is absorbed into the $E_6$, leaving $\overline{\eta}^1 - \overline{\eta}^2$ and $\overline{\eta}^1 + \overline{\eta}^2 - 2\overline{\eta}^3$
as generators of extra $U(1)$ charges, along with the 3 extra $U(1)$'s generated by the complex 
$\overline{y}^I = \overline{y}^1 + i \overline{y}^2$, 
$\overline{y}^{II} = \overline{y}^3 + i \overline{y}^4$, and
$\overline{y}^{III} = \overline{y}^5 + i \overline{y}^6$.

Instead of producing 8 copies of non-chiral generations of $SO(10)$ $\mbf{16}$'s in each $\bb_i$ sector, this NAHE variation produces 1 non-chiral generation of $E_6$ $\mbf{27}$'s in each $\{\bone,\bb_i\}$ sector set and an additional 4 non-chiral generations in each of the three $\{\bS + \bb_i + \bb_j\}$, $i\neq j$ sector. (See Appendix A.) Thus, this model corresponds to $h^{1,1} = h^{2,1} = 15$. This model
thus has the Hodge numbers and twisted sector matter distributions of the orbifold models 
$(1-2)$ and $(1-8)$ of \cite{rd1} and may be the free fermionic equivalent of one of these. 

The NAHE variation also contains 45 pairs of vector-like non-Abelian matter singlets (carrying $U(1)$ charges) with 9 pairs coming from the untwisted sector and 12 pairs from each of the three $\bb_i + \bb_j$ sectors. The untwisted sector also contains 6 copies of $\mbf{22}$ reps of the hidden sector $SO(22)$, while each $\bS + \bb_i + \bb_i$ sector produces an additional 8 copies of $\mbf{22}$ reps of $SO(22)$. The third order components of the model's superpotential are given in Appendix B. (The next lowest order terms are fifth order--there are no fourth order terms.)

In concluding this section, we note that our NAHE variation has connection with another variation discussed in \cite{cnfv} that is formed from 6 basis vectors. In that model, the sector formed by the sum of the three $\bb_i$ in
our above variation was denoted as $ ``X"$ and was added to the NAHE group. The observable sector GUT gauge group was also raised to $E_6$, with the same $U(1)$ enhancing $SO(10)$ to $E_6$. The total gauge group became 
$E_6\otimes U(1)^2 \otimes SO(4)^3 \otimes E_8$, in contrast to our
$E_6\otimes U(1)^5 \otimes SO(22)$.

\section{Examination of Example NAHE-Variation Based Models}\label{sec:disc}

In \cite{rch} we introduced a general algorithm for systematic generation of the complete set of WCFFHS gauge group models up to a chosen layer $L$ (number of basic vectors) and order $N$ (the lowest common multiple of the orders $N_i$ of the respective basis vectors $\bV_i$, whereby $N_i$ is the smallest positive integer such that $N_i \bV_i= \vec{0}\, ({\rm mod}\, 2$). (By gauge models, we mean those containing basis vectors with anti-periodic left-moving fermions.)  We have generalized our algorithm for systematical generation of models containing twisted matter sectors and, relatedly, have begun a systematic investigation of $SO(10)$ NAHE-based models \cite{rch4}. Now, with the construction of the $E_6$ NAHE-variation presented herein, we are also initiating a parallel systematic investigation of models with the NAHE-variation as their core. The general phenomenology of this new class of models, and the particular characteristics of sub-classes of models defined by their observable gauge group will be presented in an upcoming series of papers. 

One aspect of the NAHE-variation class of models that we will pursue are mirror models. That is, models with matching observable and hidden sector gauge groups and matter states. The possibility of NAHE-based mirror models was explored in \cite{nahemirror}. It was shown that GSO constraints imposed by the observable sector on the charges of hidden sector states significantly hinder realization of mirror models since the charges of observable sector states in NAHE-based models are spread out beyond half (22) of the total number of right-moving complex fermions. In fact, in \cite{nahemirror}, we showed that in a large class (perhaps all) of NAHE-based models with mirror basis vectors, these GSO constraints enforce spontaneous breaking of an initial mirror symmetry of gauge groups. 

However, our variation on the NAHE set appears more conducive to mirror model construction, since the 
$\IZ_2 \otimes \IZ_2$ twist in the NAHE variation allow observable sector states to carry charges within just the first 11 of the 22 right-moving complex fermions, allowing the additional 11 charges to be reserved for hidden sector states. Specifically, an additional three sectors denoted  $\bb^{'}_{i=1,\, 2,\, 3}$ mirroring $\bb_{i=1,\, 2,\, 3}$ in the hidden sector might be added to our NAHE variation to generate an $(E_6\otimes U(1)^5)_{\rm obs}\otimes ({E_6 \otimes U(1)^5})_{\rm hid}$ model with matching matter states.\footnote{Nevertheless singlet states carrying both observable $U(1)^5_{\rm obs}$ and
hidden $U(1)^5_{\rm hid}$ charges are likely to exist, and therefore mix the observable and hidden sectors at high orders in the superpotential.}

It should be noted that, nevertheless, the GSO projections between observable and hidden massless matter sectors can never be totally independent, since the observable and hidden matter sectors will always have a periodic complex spacetime fermion in common. Modular invariance constraints require that any pair of order-2 mirror matter sectors have at least one more non-zero complex fermion boundary condition in common, albeit the complex fermion can be either a left-moving or right-moving. Hence, for order-2 the modular invariance rules cannot be satisfied by simply adding an additional set of hidden sector mirror matter sectors $\mb^{'}_i$, for $i=1$ to $3$, with real right-moving components defined by $(\mb^{'}_i)^{n} = (\mb_i)^{44-n}$. In this case, while $\mb^{'}_i\cdot \mb_i$ satisfy modular invariance requirements, $\mb^{'}_i\cdot \mb_{j\neq i}$ do not. As we will show in \cite{gprc}, for higher order basis vectors, this requirement is lifted -- mirror observable/hidden matter sectors with either only a periodic spacetime boundary condition in common or else only a periodic spacetime and left-moving complex fermion $x$ boundary condition in common are consistent with modular invariance. 

%\newpage

Results of our full exploration of gauge and matter mirror models based on our NAHE variation will appear in \cite{gprc}. Rather than discuss the range now, we will discuss a few interesting mirror models which we have constructed by hand, randomly, and systematically.\\

As a first example, we present an interesting NAHE variation-based example of a gauge (but not matter) mirror model that satisfies modular invariance requirements. The observable and hidden sector matter basis vectors are not completely mirrors among the $\{ \overline{\eta}^{(')},\overline{y}^{(')},\overline{w}^{(')}\}$. Hence
observable and hidden sector matter are not mirror images.
The gauge group is 
($E_6)_{\rm obs} \otimes (U(1))^7 \otimes SU(4) \otimes (E_6)_{\rm hid}$.
The model is chiral with 21 $\mbf{27}$ reps and 3 $\overline{\mbf{27}}$ reps of $(E_6)_{\rm obs}$. (The untwisted sector provides 3 $\mbf{27}$'s and 3 $\overline{\mbf{27}}$'s -- the 18 net chiral reps are all from the twisted sectors.)
The model also contains 12 ${\mbf{4}}$ and 12 $\overline{\mbf{4}}$ reps (not in vector-like pairs) of $SU(4)$ and 48 $U(1)^5$ - charged non-Abelian singlets.
There are neither $\mbf{27}$ nor $\overline{\mbf{27}}$ reps of $(E_{6})_{\rm hid}$.
A net $\IZ_6$ twist from additional sectors is needed to simultaneously (1) reduce $(E_6)_{\rm obs}$ to a (semi-)GUT that does not require adjoint or higher scalar reps to induce a spontaneous symmetry breaking to the MSSM at low energy, and (2) reduce the number of copies of each matter generation from 6 to 1.
The basis vectors and GSO projection matrix are given in Tables 3a and 3b.  
\newpage

{{\bf Table 3a. NAHE Variation Mirrored Gauge Group Model Basis Vectors}}
{\large
{
\beqn
 &&\begin{tabular}{||c|c||c|c|c|c|c|c|c|c|c|c|c|c||}
 \hline \hline
 \textbf{Sec}&\textbf{N}& $\psi^\mu$ & ${x^{12}}$ & ${x^{34}}$ & ${x^{56}}$ &
        $\overline{\psi}^{1,...,5} $ &
        {$\overline{\eta}^1$}&
        {$\overline{\eta}^2$}&
        {$\overline{\eta}^3$}&
        {$\overline{\eta^{'}}^1$}&
        {$\overline{\eta^{'}}^2$}&
        {$\overline{\eta^{'}}^3$}&
        $\overline{\psi^{'}}^{1,...,5} $ \\
\hline
\hline
    {$\bone$}&2 &  1 & 1&1&1 & 1,...,1 & 1 & 1 & 1 & 1 & 1 & 1& 1,...,1 \\
    \hline
        $\mS$ &2&  1 & 1&1&1 & 0,...,0 & 0 & 0 & 0 & 0 & 0 & 0& 0,...,0 \\
\hline
  {${\mb}_1$} &2&  1 & 1&0&0 & 1,...,1 & 1 & 0 & 0 & 0 & 0 & 0& 0,...,0 \\
  \hline
  {${\mb}_2$} &2&  1 & 0&1&0 & 1,...,1 & 0 & 1 & 0 & 0 & 0 & 0& 0,...,0 \\
  \hline
  {${\mb}_3$} &2&  1 & 0&0&1 & 1,...,1 & 0 & 0 & 1 & 0 & 0 & 0& 0,...,0 \\
  \hline
{${\mb}^{'}_1$}&2& 1 & 1&0&0 & 0,...,0 & 0 & 1 & 1 & 1 & 0 & 0& 1,...,1 \\
\hline
{${\mb}^{'}_2$}&2& 1 & 0&1&0 & 0,...,0 & 1 & 0 & 1 & 0 & 1 & 0& 1,...,1 \\
\hline
{${\mb}^{'}_3$}&2& 1 & 0&0&1 & 0,...,0 & 1 & 1 & 0 & 0 & 0 & 1& 1,...,1 \\
\hline \hline
\end{tabular}
   \nonumber\\
   ~  &&  ~ \nonumber\\
   &&\begin{tabular}{||c|c||c|c|c|c|c|c|c|c|c|c|c|c||}
   \hline \hline
 \textbf{Sec}&\textbf{N}&     {$y^{1,2}$}         &
        {${\overline y}^{1,2}$}  &
        {$y^{3,4}$}         &
        {${\overline y}^{3,4}$}  &
        {$y^{5,6}$}         &
        {${\overline y}^{5,6}$}  &
        {$w^{1,2}$}         &
        {${\overline w}^{1,2}$}  &
        {$w^{3,4}$}         &
        {${\overline w}^{3,4}$}  &
        {$w^{5,6}$}         &
        {${\overline w}^{5,6}$}  \\
\hline
\hline
  {$\bone$}&2 & 1,1 & 1,1 & 1,1 & 1,1 & 1,1 & 1,1 
            & 1,1 & 1,1 & 1,1 & 1,1 & 1,1 & 1,1\\
            \hline
   $\mS$ &2   & 0,0 & 0,0 & 0,0 & 0,0 & 0,0 & 0,0 
            & 0,0 & 0,0 & 0,0 & 0,0 & 0,0 & 0,0\\
\hline
{${\mb}_1$}&2 & 0,0 & 0,0 & 1,1 & 1,1 & 1,1 & 1,1 
            & 0,0 & 0,0 & 0,0 & 0,0 & 0,0 & 0,0\\
            \hline
{${\mb}_2$}&2 & 1,1 & 1,1 & 0,0 & 0,0 & 1,1 & 1,1 
            & 0,0 & 0,0 & 0,0 & 0,0 & 0,0 & 0,0\\
            \hline
{${\mb}_3$}&2 & 1,1 & 1,1 & 1,1 & 1,1 & 0,0 & 0,0 
            & 0,0 & 0,0 & 0,0 & 0,0 & 0,0 & 0,0\\
            \hline
{${\mb^{'}}_1$}&2& 0,0 & 0,0 & 0,0 & 0,0 & 0,0 & 0,0 
            & 0,0 & 0,0 & 1,1 & 0,0 & 1,1 & 0,0\\
            \hline
{${\mb^{'}}_2$}&2& 0,0 & 0,0 & 0,0 & 0,0 & 0,0 & 0,0 
            & 1,1 & 0,0 & 0,0 & 0,0 & 1,1 & 0,0\\
            \hline
{${\mb^{'}}_3$}&2& 0,0 & 0,0 & 0,0 & 0,0 & 0,0 & 0,0 
            & 1,1 & 0,0 & 1,1 & 0,0 & 0,0 & 0,0 \\
\hline \hline
\end{tabular}
\nonumber
\eeqn
}}

{{\bf Table 3b. NAHE Variation Mirrored Gauge Group Model GSO Projection Matrix}}
{\large
{
\beqn
\left(
\begin{tabular}{c|cccccccc}
   $k_{i,j}$& $\bone$ & $\mS$ & ${\mb}_1$ & ${\mb}_2$ & ${\mb}_3$ 
                              & ${\mb}^{'}_1$ & ${\mb}^{'}_2$ & ${\mb}^{'}_3$\\
\hline
   $\bone$        &0&0&1&1&1&1&1&1\\
   $\mS$          &0&0&0&0&0&0&0&0\\
   ${\mb}_1$      &1&1&1&1&1&1&0&0\\
   ${\mb}_2$      &1&1&1&1&1&0&1&0\\
   ${\mb}_3$      &1&1&1&1&1&0&0&1\\
   ${\mb}^{'}_1$  &1&1&0&0&0&1&0&0\\
   ${\mb}^{'}_2$  &1&1&0&0&0&0&1&0\\
   ${\mb}^{'}_3$  &1&1&0&0&0&0&0&1\\
\end{tabular}\right)
\nonumber
\eeqn
}}

The next two models were generated randomly during the testing phases of Baylor University's new software for generating free fermionic heterotic string models, the FF Framework. These models have semi-mirrored gauge groups and mirrored matter states, and are non-supersymmetric. (The output format for basis vectors and particle content is a standard feature of the FF Framework.)
\vspace{3 mm}\\
In Table 4.a, the first of these two models shows two gauge groups (both $SO(11)$) which have mirrored matter states. It is also worth noting that there is only one ``mixed" matter state which transforms under both. The other non-Abelian matter representations of those groups do not couple with one another. While there is an additional $SO(10)$ gauge group, the matter representations of that group also do not couple with the matter representations of the two $SO(11)$'s. The mirrored gauge and matter representations of this model are not readily apparent from the basis vectors, as the non-simply laced gauge groups in the model are produced by a twisted basis. The basis vectors and the $k_{ij}$ matrix are in appendix  \ref{sec:Mirror_Inputs}. The fact that the mirroring of this model is not readily apparent emphasizes the need for systematic construction of the NAHE-variation extensions, as is currently in progress. Without a complete search of all possible basis vectors, models in which this symmetry is present may go unnoticed.
\setlength\extrarowheight{1.15 pt}
\begin{flushleft}
\textbf{Table 3.d - Mirrored Matter Representations Model 2 - Particle Content}
\vspace{3 mm}\\
\begin{tabular}{||c||c|c|c||}
\hline
\hline
\textbf{QTY}&$SO(11)$&$SO(11)$&$SO(10)$\\
\hline
\hline
8&$1$&$1$&$\overline{16}$\\ 
\hline
88&$1$&$1$&$1$\\
\hline
12&$1$&$1$&$10$\\
\hline
12&$1$&$1$&$16$\\
\hline
16&$1$&$11$&$1$\\
\hline
2&$1$&$32$&$1$\\
\hline
16&$11$&$1$&$1$\\
\hline
1&$11$&$11$&$1$\\
\hline
2&$32$&$1$&$1$\\
\hline
\hline
\end{tabular}

\vspace{3 mm}
\textbf{Total Matter Representations: 157}
\vspace{3 mm}\\
\textbf{Number of ST SUSYs: 0}
\vspace {6 mm}\\

The next model, presented in table 3.b, has mirrored $SO(10)$ gauge groups. In this model, the matter representations of these gauge groups do not couple with one another, but each $SO(10)$ has a single state which couples to the additional, non-mirrored gauge group $SO(14)$. In this case, the mirroring of the gauge groups is evident from the basis vectors, shown in table 3.c. In particular, the elements $\bar{\psi}^{1,1*,...,5,5*}~\bar{\eta}^{1,1*,...,3,3*}$ and the elements $\bar{\phi}^{1,1*,...,8,8*}$ are exactly the same. The real elements $\bar{y}^{1,...,6}~\bar{w}^{1,...,6}$ can also be mirrored exactly. The $k_{ij}$ matrix is in appendix \ref{sec:Mirror_Inputs}.\\ \vspace{3 mm}
\newpage
\textbf{Table 3.e - Mirror Matter Representations Model 3 - Particle Content}
\vspace{3 mm}\\
\begin{tabular}{||c||c|c|c||}
\hline
\hline
\textbf{QTY}&$SO(10)$&$SO(10)$&$SO(14)$\\
\hline
\hline
8&$\overline{16}$&$1$&$1$\\
\hline
8&$1$&$\overline{16}$&$1$\\
\hline
40&$1$&$1$&$1$\\
\hline
12&$1$&$1$&$14$\\
\hline
14&$1$&$10$&$1$\\
\hline
1&$1$&$10$&$14$\\
\hline
8&$1$&$16$&$1$\\
\hline
14&$10$&$1$&$1$\\
\hline
1&$10$&$1$&$14$\\
\hline
8&$16$&$1$&$1$\\
\hline
\hline
\end{tabular}

\vspace{3 mm}
\textbf{Total Matter Representations: 114}
\vspace{3 mm}\\
\textbf{Number of ST SUSYs: 0}
\vspace{6 mm}\\

\textbf{Table 3.f - Mirror Matter Representations Model 3 - Basis Vectors}
\vspace{3 mm}\\
\begin{tabular}{||c|c||c|c||c|c|c||c|c|c||}
\hline
\hline
\textbf{Sec}&\textbf{N}&$\psi^1$&$\psi^{1*}$&$x^1$
&$y^1$
&$w^1$
&$x^2$
&$y^2$
&$w^2$\\
\hline
1&2&1&1&1&1&1&1&1&1\\
\hline
$b_1$&2&1&1&1&0&0&1&0&0\\
\hline
$b_2$&2&1&1&0&1&0&0&1&0\\
\hline
$b_3$&2&1&1&0&1&0&0&1&0\\
\hline
$\alpha_{0}$&3&1&1&1&0&0&1&0&0\\
\hline
\hline
\end{tabular}

\vspace{3 mm}
\begin{tabular}{||c|c||c|c|c||c|c|c||c|c|c||c|c|c||}
\hline
\hline
\textbf{Sec}&\textbf{N}&$x^3$&$y^3$&$w^3$&$x^4$&$y^4$&$w^4$&$x^5$&$y^5$&$w^5$&$x^6$&$y^6$&$w^6$\\
\hline
1&2&1&1&1&1&1&1&1&1&1&1&1&1\\
\hline
$b_1$&2&0&1&0&0&1&0&0&1&0&0&1&0\\
\hline
$b_2$&2&1&0&0&1&0&0&0&1&0&0&1&0\\
\hline
$b_3$&2&0&1&0&0&1&0&1&0&0&1&0&0\\
\hline
$\alpha_{0}$&3&1&0&0&1&0&0&1&0&0&1&0&0\\
\hline
\hline
\end{tabular}

\vspace{3 mm}

\begin{tabular}{||c|c||c|c|c|c|c|c|c|c|c|c||c|c|c|c|c|c||}
\hline
\hline
\textbf{Sec}&\textbf{N}&$\overline{\psi}^{1}$&$\overline{\psi}^{1*}$&$\overline{\psi}^{2}$&$\overline{\psi}^{2*}$&$\overline{\psi}^{3}$&$\overline{\psi}^{3*}$&$\overline{\psi}^{4}$&$\overline{\psi}^{4*}$&$\overline{\psi}^{5}$&$\overline{\psi}^{5*}$&$\overline{\eta}^{1}$&$\overline{\eta}^{1*}$&$\overline{\eta}^{2}$&$\overline{\eta}^{2*}$&$\overline{\eta}^{3}$&$\overline{\eta}^{3*}$\\
\hline
1&2&1&1&1&1&1&1&1&1&1&1&1&1&1&1&1&1\\
\hline
$b_1$&2&1&1&1&1&1&1&1&1&1&1&1&1&0&0&0&0\\
\hline
$b_2$&2&1&1&1&1&1&1&1&1&1&1&0&0&1&1&0&0\\
\hline
$b_3$&2&1&1&1&1&1&1&1&1&1&1&0&0&0&0&1&1\\
\hline
$\alpha_{0}$&3&0&0&0&0&0&0&0&0&0&0&0&0&1&1&1&1\\
\hline
\end{tabular}

\vspace{3 mm}
\begin{tabular}{||c|c||c|c|c|c|c|c||c|c|c|c|c|c||}
\hline
\hline
\textbf{Sec}&\textbf{N}&$\overline{y}^{1}$&$\overline{y}^{2}$&$\overline{y}^{3}$&$\overline{y}^{4}$&$\overline{y}^{5}$&$\overline{y}^{6}$&$\overline{w}^{1}$&$\overline{w}^{2}$&$\overline{w}^{3}$&$\overline{w}^{4}$&$\overline{w}^{5}$&$\overline{w}^{6}$\\
\hline
1&2&1&1&1&1&1&1&1&1&1&1&1&1\\
\hline
$b_1$&2&0&0&1&1&1&1&0&0&0&0&0&0\\
\hline
$b_2$&2&1&1&0&0&1&1&0&0&0&0&0&0\\
\hline
$b_3$&2&1&1&1&1&0&0&0&0&0&0&0&0\\
\hline
$\alpha_{0}$&3&2&2&2&2&2&2&2&0&1&2&0&1\\
\hline
\end{tabular}

\vspace{3 mm}
\begin{tabular}{||c|c||c|c|c|c|c|c|c|c|c|c|c|c|c|c|c|c||}
\hline
\hline
\textbf{Sec}&\textbf{N}&$\overline{\phi}^{1}$&$\overline{\phi}^{1*}$&$\overline{\phi}^{2}$&$\overline{\phi}^{2*}$&$\overline{\phi}^{3}$&$\overline{\phi}^{3*}$&$\overline{\phi}^{4}$&$\overline{\phi}^{4*}$&$\overline{\phi}^{5}$&$\overline{\phi}^{5*}$&$\overline{\phi}^{6}$&$\overline{\phi}^{6*}$&$\overline{\phi}^{7}$&$\overline{\phi}^{7*}$&$\overline{\phi}^{8}$&$\overline{\phi}^{8*}$\\
\hline
1&2&1&1&1&1&1&1&1&1&1&1&1&1&1&1&1&1\\
\hline
$b_1$&2&0&0&0&0&0&0&0&0&0&0&0&0&0&0&0&0\\
\hline
$b_2$&2&0&0&0&0&0&0&0&0&0&0&0&0&0&0&0&0\\
\hline
$b_3$&2&0&0&0&0&0&0&0&0&0&0&0&0&0&0&0&0\\
\hline
$\alpha_{0}$&3&0&0&0&0&0&0&0&0&0&0&0&0&1&1&1&1\\
\hline
\end{tabular}

\vspace{6 mm}
The last (fourth) is a mirrored with regard to both gauge groups and matter representations. The observable and hidden gauge groups are $(E_{6})_{obs}\otimes (E_{6})_{hid}$ with an additional $SO(14)$. The matter representations from each gauge group do not interact under gauge transformations with any other group. The $E_6$ representations also do not display chirality. This model was found during a systematic investigation of the NAHE-variation basis vectors, with only one additional basis vector of order 3. A full report on that investigation will be presented soon. The basis vectors which produced this model are presented in appendix \ref{sec:Mirror_Inputs}.\\
\vspace {5 mm}
\textbf{Table 3.g Mirror Matter Representation Model 4 - Particle Content}
\vspace {2 mm}\\
\begin{tabular}{||c||c|c|c||}
\hline
\hline
\textbf{QTY}&$SO(14)$&$E_6$&$E_6$\\
\hline
\hline
6&$1$&$\overline{27}$&$1$\\
\hline
6&$1$&$1$&$\overline{27}$\\
\hline
6&$1$&$1$&$27$\\
\hline
6&$1$&$27$&$1$\\
\hline
12&$14$&$1$&$1$\\
\hline
\hline
\end{tabular}
\vspace{3 mm}
\\
\textbf{Total Matter Representations: 36}
\vspace{3 mm}
\\
\textbf{Number of ST SUSYs: 2}
\vspace {6 mm}\\

\end{flushleft}
\section{Conclusion}
We introduced a variation on the traditional NAHE set with the intent of creating a new class of models for examination. These models are more easily open to geometric interpretation, and can thus be compared to other heterotic string construction methods, specifically the orbifold method. Systematic construction and examination of the overlap between orbifolds and free fermions is crucial to a better understanding of the string theory landscape.\\

Additionally, models based on this variation are more apt to display a mirroring between the hidden and observable gauge groups, due to having larger sets of matching boundary conditions for the right moving part of the basis set. This was demonstrated by the models presented herein. It was also shown that, although this mirroring can be evident in the basis vectors which produce the model, it is not always so. Thus, a complete examination of the input space for this class of models is needed to fully explore these ``mirror models." Such an investigation is currently underway.
\section{Acknowledgements}

Research funding leading to this manuscript was partially provided by Baylor URC grant 0301533BP.

\newpage
\appendix
{
\section{NAHE Variation}
\subsection{$E_6 \otimes U(1)^5 \otimes SO(22)$ States}

\noindent Note: all $U(1)$ charges below have been multiplied by a factor or 4 to eliminate fractions.

\noindent
\tablehead{\hline \hline
HWS Sector& State &  $E_6$ & $U(1)_1$ & $U(1)_2$ & $U(1)_3$ & $U(1)_4$ & $U(1)_5$ & $SO(22)$\\ \hline \hline }
\tabletail{\hline} %9 columns per line.
\tablelasttail{\hline \hline}
\begin{supertabular}{|l|l||r|rrrrr|r|}
$\bone$ 
    &$G_1$                  & $\mbf{27}$&   0&    8&    0&    0&    0& $\mbf 1$\\
    &$G_2$                  & $\mbf{27}$&   4&   -4&    0&    0&    0& $\mbf 1$\\
    &$G_3$                  & $\mbf{27}$&  -4&   -4&    0&    0&    0& $\mbf 1$\\
    &$\oG_1$             & $\ol{\mbf{27}}$&  0&   -8&    0&    0&    0& $\mbf 1$\\
    &$\oG_2$             & $\ol{\mbf{27}}$& -4&    4&    0&    0&    0& $\mbf 1$\\
    &$\oG_3$             & $\ol{\mbf{27}}$&  4&    4&    0&    0&    0& $\mbf 1$\\
\hline
$\bS+\bb_1+\bb_2$
    &$G_4$                  & $\mbf{27}$&   0&   -4&   -2&   -2&    0& $\mbf 1$\\
    &$G_5$                  & $\mbf{27}$&   0&   -4&   -2&    2&    0& $\mbf 1$\\
    &$G_6$                  & $\mbf{27}$&   0&   -4&    2&   -2&    0& $\mbf 1$\\
    &$G_7$                  & $\mbf{27}$&   0&   -4&    2&    2&    0& $\mbf 1$\\
    &$\oG_4$            & $\ol{\mbf{27}}$&   0&    4&    2&    2&    0& $\mbf 1$\\
    &$\oG_5$            & $\ol{\mbf{27}}$&   0&    4&    2&   -2&    0& $\mbf 1$\\
    &$\oG_6$            & $\ol{\mbf{27}}$&   0&    4&   -2&    2&    0& $\mbf 1$\\
    &$\oG_7$            & $\ol{\mbf{27}}$&   0&    4&   -2&   -2&    0& $\mbf 1$\\
\hline
$\bS+\bb_1+\bb_3$ 
     &$G_8$                  & $\mbf{27}$&  -2&    2&   -2&    0&   -2& $\mbf 1$\\
     &$G_9$                  & $\mbf{27}$&  -2&    2&   -2&    0&    2& $\mbf 1$\\
     &$G_{10}$                  & $\mbf{27}$&  -2&    2&    2&    0&   -2& $\mbf 1$\\
     &$G_{11}$                  & $\mbf{27}$&  -2&    2&    2&    0&    2& $\mbf 1$\\
     &$\oG_8$            & $\ol{\mbf{27}}$&   2&   -2&    2&    0&    2& $\mbf 1$\\
     &$\oG_9$            & $\ol{\mbf{27}}$&   2&   -2&    2&    0&   -2& $\mbf 1$\\
     &$\oG_{10}$            & $\ol{\mbf{27}}$&   2&   -2&   -2&    0&    2& $\mbf 1$\\
     &$\oG_{11}$            & $\ol{\mbf{27}}$&   2&   -2&   -2&    0&   -2& $\mbf 1$\\
\hline
$\bS+\bb_2+\bb_3$  
    &$G_{12}$                  & $\mbf{27}$&   2&    2&    0&   -2&   -2& $\mbf 1$\\
    &$G_{13}$                  & $\mbf{27}$&   2&    2&    0&   -2&    2& $\mbf 1$\\
    &$G_{14}$                  & $\mbf{27}$&   2&    2&    0&    2&   -2& $\mbf 1$\\
    &$G_{15}$                  & $\mbf{27}$&   2&    2&    0&    2&    2& $\mbf 1$\\
    &$\oG_{12}$            & $\ol{\mbf{27}}$&  -2&   -2&    0&    2&    2& $\mbf 1$\\
    &$\oG_{13}$            & $\ol{\mbf{27}}$&  -2&   -2&    0&    2&   -2& $\mbf 1$\\
    &$\oG_{14}$            & $\ol{\mbf{27}}$&  -2&   -2&    0&   -2&    2& $\mbf 1$\\
    &$\oG_{15}$            & $\ol{\mbf{27}}$&  -2&   -2&    0&   -2&   -2& $\mbf 1$\\
\hline             
$\bone$ 
    &$\p_{1}$ ($\op_{1}$)                  & $\mbf 1$&    0&    0&    0&   -4&   -4&            $\mbf 1$\\
    &$\p_{2}$ ($\op_{2}$)                  & $\mbf 1$&    0&    0&    0&   -4&    4&            $\mbf 1$\\
    &$\p_{3}$ ($\op_{3}$)                  & $\mbf 1$&    0&    0&   -4&    0&   -4&            $\mbf 1$\\
    &$\p_{4}$ ($\op_{4}$)                  & $\mbf 1$&    0&    0&   -4&    0&    4&            $\mbf 1$\\
    &$\p_{5}$ ($\op_{5}$)                  & $\mbf 1$&    0&    0&   -4&   -4&    0&            $\mbf 1$\\
    &$\p_{6}$ ($\op_{6}$)                  & $\mbf 1$&    0&    0&   -4&    4&    0&            $\mbf 1$\\
    &$\p_{7}$ ($\op_{7}$)                  & $\mbf 1$&    4&  -12&    0&    0&    0&            $\mbf 1$\\
    &$\p_{8}$ ($\op_{8}$)                  & $\mbf 1$&    4&   12&    0&    0&    0&            $\mbf 1$\\
    &$\p_{9}$ ($\op_{9}$)                  & $\mbf 1$&   -8&    0&    0&    0&    0&            $\mbf 1$\\
\hline
$\bS+\bb_1+\bb_2$  
    &$\s_{1}$ ($\bs_{1}$)                  & $\mbf 1$&    0&   12&    2&    2&    0&            $\mbf 1$\\
    &$\s_2$ ($\bs_{2}$)                  & $\mbf 1$&    0&   12&    2&   -2&    0&            $\mbf 1$\\
    &$\s_3$ ($\bs_{3}$)                  & $\mbf 1$&    0&   12&   -2&    2&    0&            $\mbf 1$\\
    &$\s_4$ ($\bs_{4}$)                  & $\mbf 1$&    0&   12&   -2&   -2&    0&            $\mbf 1$\\
    &$\s_5$ ($\bs_{5}$)                  & $\mbf 1$&    4&    0&    2&    2&   -4&            $\mbf 1$\\
    &$\s_6$ ($\bs_{6}$)                  & $\mbf 1$&    4&    0&    2&    2&    4&            $\mbf 1$\\
    &$\s_7$ ($\bs_{7}$)                  & $\mbf 1$&    4&    0&    2&   -2&   -4&            $\mbf 1$\\
    &$\s_8$ ($\bs_{8}$)                  & $\mbf 1$&    4&    0&    2&   -2&    4&            $\mbf 1$\\
    &$\s_9$ ($\bs_{9}$)                  & $\mbf 1$&    4&    0&   -2&    2&   -4&            $\mbf 1$\\
    &$\s_{10}$ ($\bs_{10}$)                 & $\mbf 1$&    4&    0&   -2&    2&    4&            $\mbf 1$\\
    &$\s_{11}$ ($\bs_{11}$)                 & $\mbf 1$&    4&    0&   -2&   -2&   -4&            $\mbf 1$\\
    &$\s_{12}$ ($\bs_{12}$)                 & $\mbf 1$&    4&    0&   -2&   -2&    4&            $\mbf 1$\\
\hline
$\bS+\bb_1+\bb_3$ 
     &$\s_{13}$ ($\bs_{13}$)                 & $\mbf 1$&    2&    6&    2&   -4&    2&            $\mbf 1$\\
     &$\s_{14}$ ($\bs_{14}$)                 & $\mbf 1$&    2&    6&    2&   -4&   -2&            $\mbf 1$\\
     &$\s_{15}$ ($\bs_{15}$)                 & $\mbf 1$&    2&    6&    2&    4&    2&            $\mbf 1$\\
     &$\s_{16}$ ($\bs_{16}$)                 & $\mbf 1$&    2&    6&    2&    4&   -2&            $\mbf 1$\\
     &$\s_{17}$ ($\bs_{17}$)                 & $\mbf 1$&    2&    6&   -2&   -4&    2&            $\mbf 1$\\
     &$\s_{18}$ ($\bs_{18}$)                 & $\mbf 1$&    2&    6&   -2&   -4&   -2&            $\mbf 1$\\
     &$\s_{19}$ ($\bs_{19}$)                 & $\mbf 1$&    2&    6&   -2&    4&    2&            $\mbf 1$\\
     &$\s_{20}$ ($\bs_{20}$)                 & $\mbf 1$&    2&    6&   -2&    4&   -2&            $\mbf 1$\\
     &$\s_{21}$ ($\bs_{21}$)                 & $\mbf 1$&    6&   -6&    2&    0&    2&            $\mbf 1$\\
     &$\s_{22}$ ($\bs_{22}$)                 & $\mbf 1$&    6&   -6&    2&    0&   -2&            $\mbf 1$\\
     &$\s_{23}$ ($\bs_{23}$)                 & $\mbf 1$&    6&   -6&   -2&    0&    2&            $\mbf 1$\\
     &$\s_{24}$ ($\bs_{24}$)                 & $\mbf 1$&    6&   -6&   -2&    0&   -2&            $\mbf 1$\\  
\hline
$\bS+\bb_2+\bb_3$
    &$\s_{25}$  ($\bs_{25}$)                & $\mbf 1$&   -2&    6&   -4&    2&    2&            $\mbf 1$\\
    &$\s_{26}$  ($\bs_{26}$)                & $\mbf 1$&   -2&    6&   -4&    2&   -2&            $\mbf 1$\\
    &$\s_{27}$  ($\bs_{27}$)                & $\mbf 1$&   -2&    6&   -4&   -2&    2&            $\mbf 1$\\
    &$\s_{28}$  ($\bs_{28}$)                & $\mbf 1$&   -2&    6&   -4&   -2&   -2&            $\mbf 1$\\
    &$\s_{29}$  ($\bs_{29}$)                & $\mbf 1$&   -2&    6&    4&    2&    2&            $\mbf 1$\\
    &$\s_{30}$  ($\bs_{30}$)                & $\mbf 1$&   -2&    6&    4&    2&   -2&            $\mbf 1$\\
    &$\s_{31}$  ($\bs_{31}$)                & $\mbf 1$&   -2&    6&    4&   -2&    2&            $\mbf 1$\\
    &$\s_{32}$  ($\bs_{32}$)                & $\mbf 1$&   -2&    6&    4&   -2&   -2&            $\mbf 1$\\
    &$\s_{33}$  ($\bs_{33}$)                & $\mbf 1$&   -6&   -6&    0&    2&    2&            $\mbf 1$\\
    &$\s_{34}$  ($\bs_{34}$)                & $\mbf 1$&   -6&   -6&    0&    2&   -2&            $\mbf 1$\\
    &$\s_{35}$  ($\bs_{35}$)                & $\mbf 1$&   -6&   -6&    0&   -2&    2&            $\mbf 1$\\
    &$\s_{36}$  ($\bs_{36}$)                & $\mbf 1$&   -6&   -6&    0&   -2&   -2&            $\mbf 1$\\
\hline 
$\bone$
    &$H_1$                  & $\mbf 1$&    0&    0&    0&    0&   -4& $\mbf{22}$\\
    &$H_2$                  & $\mbf 1$&    0&    0&    0&    0&    4& $\mbf{22}$\\
    &$H_3$                  & $\mbf 1$&    0&    0&    0&   -4&    0& $\mbf{22}$\\
    &$H_4$                  & $\mbf 1$&    0&    0&    0&    4&    0& $\mbf{22}$\\
    &$H_5$                  & $\mbf 1$&    0&    0&   -4&    0&    0& $\mbf{22}$\\
    &$H_6$                  & $\mbf 1$&    0&    0&    4&    0&    0& $\mbf{22}$\\
\hline
$\bS+\bb_1+\bb_2$
    &$H_7$                  & $\mbf 1$&    4&    0&    2&    2&    0& $\mbf{22}$\\
    &$H_8$                  & $\mbf 1$&    4&    0&    2&   -2&    0& $\mbf{22}$\\
    &$H_9$                  & $\mbf 1$&    4&    0&   -2&    2&    0& $\mbf{22}$\\
    &$H_{10}$                  & $\mbf 1$&    4&    0&   -2&   -2&    0& $\mbf{22}$\\
    &$H_{11}$                  & $\mbf 1$&   -4&    0&    2&    2&    0& $\mbf{22}$\\
    &$H_{12}$                  & $\mbf 1$&   -4&    0&    2&   -2&    0& $\mbf{22}$\\
    &$H_{13}$                  & $\mbf 1$&   -4&    0&   -2&    2&    0& $\mbf{22}$\\
    &$H_{14}$                  & $\mbf 1$&   -4&    0&   -2&   -2&    0& $\mbf{22}$\\
\hline
$\bS+\bb_1+\bb_3$ 
     &$H_{15}$                  & $\mbf 1$&    2&    6&    2&    0&    2& $\mbf{22}$\\
     &$H_{16}$                  & $\mbf 1$&    2&    6&    2&    0&   -2& $\mbf{22}$\\
     &$H_{17}$                  & $\mbf 1$&    2&    6&   -2&    0&    2& $\mbf{22}$\\
     &$H_{18}$                  & $\mbf 1$&    2&    6&   -2&    0&   -2& $\mbf{22}$\\
     &$H_{19}$                  & $\mbf 1$&   -2&   -6&    2&    0&    2& $\mbf{22}$\\
     &$H_{20}$                  & $\mbf 1$&   -2&   -6&    2&    0&   -2& $\mbf{22}$\\
     &$H_{21}$                  & $\mbf 1$&   -2&   -6&   -2&    0&    2& $\mbf{22}$\\
     &$H_{22}$                  & $\mbf 1$&   -2&   -6&   -2&    0&   -2& $\mbf{22}$\\
\hline
$\bS+\bb_2+\bb_3$
    &$H_{23}$                  & $\mbf 1$&   -2&    6&    0&    2&    2& $\mbf{22}$\\
    &$H_{24}$                  & $\mbf 1$&   -2&    6&    0&    2&   -2& $\mbf{22}$\\
    &$H_{25}$                  & $\mbf 1$&   -2&    6&    0&   -2&    2& $\mbf{22}$\\
    &$H_{26}$                  & $\mbf 1$&   -2&    6&    0&   -2&   -2& $\mbf{22}$\\
    &$H_{27}$                  & $\mbf 1$&    2&   -6&    0&    2&    2& $\mbf{22}$\\
    &$H_{28}$                  & $\mbf 1$&    2&   -6&    0&    2&   -2& $\mbf{22}$\\
    &$H_{29}$                  & $\mbf 1$&    2&   -6&    0&   -2&    2& $\mbf{22}$\\
    &$H_{30}$                  & $\mbf 1$&    2&   -6&    0&   -2&   -2& $\mbf{22}$\\
\hline
\end{supertabular}
\subsection{$E_6 \otimes U(1)^5 \otimes SO(22)$ Third Order Superpotential (No Fourth Order Terms Exist)}
\tablehead{}
\tabletail{&&&&\\}
\tablelasttail{&&&&\\}
\begin{supertabular}{lllll}
\phantom{+}   $\oG_1$  $G_2$  $G_3$
& +   $G_1$ $\oG_2$  $\p_{7}$ 
& +   $G_1$ $\oG_3$ $\op_{8}$ 
& +   $G_1$  $G_4$  $G_7$
& +   $G_1$  $G_5$  $G_6$                     
 \\
+     $G_1$ $\oG_4$ $\os_{1}$ 
& +   $G_1$ $\oG_5$ $\os_{2}$ 
& +   $G_1$ $\oG_6$ $\os_{3}$ 
& +   $G_1$ $\oG_7$ $\os_{4}$ 
& +   $G_2$ $\oG_1$ $\op_{7}$                
\\
+     $G_2$ $\oG_3$  $\p_{9}$ 
& +   $G_2$  $G_8$  $G_{11}$
& +   $G_2$  $G_9$  $G_{10}$
& +   $G_2$ $\oG_8$ $\os_{21}$
& +   $G_2$ $\oG_9$ $\os_{22}$              
\\
  +   $G_2$ $\oG_{10}$ $\os_{23}$
& +   $G_2$ $\oG_{11}$ $\os_{24}$
& +   $G_3$ $\oG_1$  $\p_{8}$ 
& +   $G_3$ $\oG_2$ $\op_{9}$ 
& +   $G_3$  $G_{12}$  $G_{15}$           
  \\
  +   $G_3$  $G_{13}$  $G_{14}$
& +   $G_3$ $\oG_{12}$ $\os_{33}$
& +   $G_3$ $\oG_{13}$ $\os_{34}$
& +   $G_3$ $\oG_{14}$ $\os_{35}$
& +   $G_3$ $\oG_{15}$ $\os_{36}$         
\\
  +  $\oG_1$ $\oG_2$ $\oG_3$
& +  $\oG_1$  $G_4$  $\s_{1}$ 
& +  $\oG_1$  $G_5$  $\s_2$   
& +  $\oG_1$  $G_6$  $\s_3$ 
& +  $\oG_1$  $G_7$  $\s_4$                   
\\ 
  +  $\oG_1$ $\oG_4$ $\oG_7$
& +  $\oG_1$ $\oG_5$ $\oG_6$
& +  $\oG_2$  $G_8$  $\s_{21}$
& +  $\oG_2$  $G_9$  $\s_{22}$
& +  $\oG_2$  $G_{10}$  $\s_{23}$         
 \\
  +  $\oG_2$  $G_{11}$  $\s_{24}$
& +  $\oG_2$ $\oG_8$ $\oG_{11}$
& +  $\oG_2$ $\oG_9$ $\oG_{10}$
& +  $\oG_3$  $G_{12}$  $\s_{33}$
& +  $\oG_3$  $G_{13}$  $\s_{34}$          
 \\
  +  $\oG_3$  $G_{14}$  $\s_{35}$
& +  $\oG_3$  $G_{15}$  $\s_{36}$
& +  $\oG_3$ $\oG_{12}$ $\oG_{15}$
& +  $\oG_3$ $\oG_{13}$ $\oG_{14}$
& +   $G_4$  $G_{10}$  $G_{15}$             
 \\
  +   $G_4$  $G_{11}$  $G_{14}$
& +   $G_4$ $\oG_7$ $\op_{5}$   
& +   $G_4$ $\oG_{10}$  $\s_{30}$
& +   $G_4$ $\oG_{11}$  $\s_{29}$
& +   $G_4$ $\oG_{14}$  $\s_{16}$            
 \\
  +   $G_4$ $\oG_{15}$  $\s_{15}$
& +   $G_5$  $G_{10}$  $G_{13}$
& +   $G_5$  $G_{11}$  $G_{12}$
& +   $G_5$ $\oG_6$ $\op_{6}$ 
& +   $G_5$ $\oG_{10}$  $\s_{32}$             
 \\
  +   $G_5$ $\oG_{11}$  $\s_{31}$
& +   $G_5$ $\oG_{12}$  $\s_{14}$
& +   $G_5$ $\oG_{13}$  $\s_{13}$
& +   $G_6$  $G_8$  $G_{15}$
& +   $G_6$  $G_9$  $G_{14}$                      
 \\
  +   $G_6$ $\oG_5$  $\p_{6}$ 
& +   $G_6$ $\oG_8$  $\s_{26}$
& +   $G_6$ $\oG_9$  $\s_{25}$
& +   $G_6$ $\oG_{14}$  $\s_{20}$
& +   $G_6$ $\oG_{15}$  $\s_{19}$                 
 \\
  +   $G_7$  $G_8$  $G_{13}$
& +   $G_7$  $G_9$  $G_{12}$
& +   $G_7$ $\oG_4$  $\p_{5}$ 
& +   $G_7$ $\oG_8$  $\s_{28}$
& +   $G_7$ $\oG_9$  $\s_{27}$                   
     \\
  +   $G_7$ $\oG_{12}$  $\s_{18}$
& +   $G_7$ $\oG_{13}$  $\s_{17}$
& +   $G_8$ $\oG_6$ $\os_{28}$
& +   $G_8$ $\oG_7$ $\os_{26}$   
&                                                                              
 \\
  +   $G_8$ $\oG_{11}$ $\op_{3}$ 
& +   $G_8$ $\oG_{13}$  $\s_8$ 
& +   $G_8$ $\oG_{15}$  $\s_6$ 
& +   $G_9$ $\oG_6$ $\os_{27}$
& +   $G_9$ $\oG_7$ $\os_{25}$                    
  \\
  +   $G_9$ $\oG_{10}$ $\op_{4}$ 
& +   $G_9$ $\oG_{12}$  $\s_7$ 
& +   $G_9$ $\oG_{14}$  $\s_5$ 
& +   $G_{10}$ $\oG_4$ $\os_{32}$
& +   $G_{10}$ $\oG_5$ $\os_{30}$                
 \\
  +   $G_{10}$ $\oG_9$  $\p_{4}$ 
& +   $G_{10}$ $\oG_{13}$  $\s_{12}$
& +   $G_{10}$ $\oG_{15}$  $\s_{10}$
& +   $G_{11}$ $\oG_4$ $\os_{31}$
& +   $G_{11}$ $\oG_5$ $\os_{29}$                 
 \\
  +   $G_{11}$ $\oG_8$  $\p_{3}$ 
& +   $G_{11}$ $\oG_{12}$  $\s_{11}$
& +   $G_{11}$ $\oG_{14}$  $\s_9$ 
& +   $G_{12}$ $\oG_5$ $\os_{14}$
& +   $G_{12}$ $\oG_7$ $\os_{18}$                  
 \\
  +   $G_{12}$ $\oG_9$ $\os_{7}$
& +   $G_{12}$ $\oG_{11}$ $\os_{11}$
& +   $G_{12}$ $\oG_{15}$ $\op_{1}$ 
& +   $G_{13}$ $\oG_5$ $\os_{13}$
& +   $G_{13}$ $\oG_7$ $\os_{17}$                   
\\
  +   $G_{13}$ $\oG_8$ $\os_{8}$ 
& +   $G_{13}$ $\oG_{10}$ $\os_{12}$
& +   $G_{13}$ $\oG_{14}$ $\op_{2}$ 
& +   $G_{14}$ $\oG_4$ $\os_{16}$
& +   $G_{14}$ $\oG_6$ $\os_{20}$                   
 \\
  +   $G_{14}$ $\oG_9$ $\os_{5}$ 
& +   $G_{14}$ $\oG_{11}$ $\os_{9}$ 
& +   $G_{14}$ $\oG_{13}$  $\p_{2}$ 
& +   $G_{15}$ $\oG_4$ $\os_{15}$
& +   $G_{15}$ $\oG_6$ $\os_{19}$                    
\\
  +   $G_{15}$ $\oG_8$ $\os_{6}$ 
& +   $G_{15}$ $\oG_{10}$ $\os_{10}$
& +   $G_{15}$ $\oG_{12}$  $\p_{1}$ 
& +  $\oG_4$ $\oG_{10}$ $\oG_{15}$
& +  $\oG_4$ $\oG_{11}$ $\oG_{14}$                  
\\
  +  $\oG_5$ $\oG_{10}$ $\oG_{13}$
& +  $\oG_5$ $\oG_{11}$ $\oG_{12}$
& +  $\oG_6$ $\oG_8$ $\oG_{15}$
& +  $\oG_6$ $\oG_9$ $\oG_{14}$
& +  $\oG_7$ $\oG_8$ $\oG_{13}$                       
\\
  +  $\oG_7$ $\oG_9$ $\oG_{12}$
& +   $\p_{1}$   $\p_{4}$  $\op_{5}$ 
& +   $\p_{1}$   $\p_{6}$  $\op_{3}$ 
& +   $\p_{1}$   $\s_{25}$ $\os_{26}$
& +   $\p_{1}$   $\s_{29}$ $\os_{30}$                      
 \\
  +   $\p_{1}$   $\s_{33}$ $\os_{36}$
& +   $\p_{1}$   $H_2$  $H_4$
& +   $\p_{1}$   $H_{23}$  $H_{27}$
& +   $\p_{2}$   $\p_{3}$  $\op_{5}$ 
& +   $\p_{2}$   $\p_{6}$  $\op_{4}$                         
 \\ 
  +   $\p_{2}$   $\s_{26}$ $\os_{25}$
& +   $\p_{2}$   $\s_{30}$ $\os_{29}$
& +   $\p_{2}$   $\s_{34}$ $\os_{35}$
& +   $\p_{2}$   $H_1$  $H_4$
& +   $\p_{2}$   $H_{24}$  $H_{28}$                      
   \\
  +   $\p_{3}$  $\op_{1}$  $\op_{6}$ 
& +   $\p_{3}$   $\s_{13}$ $\os_{18}$
& +   $\p_{3}$   $\s_{15}$ $\os_{20}$
& +   $\p_{3}$   $\s_{21}$ $\os_{24}$
& +   $\p_{3}$   $H_2$  $H_6$                                     
\\
  +   $\p_{3}$   $H_{15}$  $H_{19}$
& +   $\p_{4}$  $\op_{2}$  $\op_{6}$ 
& +   $\p_{4}$   $\s_{14}$ $\os_{17}$
& +   $\p_{4}$   $\s_{16}$ $\os_{19}$
& +   $\p_{4}$   $\s_{22}$ $\os_{23}$                         
\\
  +   $\p_{4}$   $H_1$  $H_6$
& +   $\p_{4}$   $H_{16}$  $H_{20}$
& +   $\p_{5}$  $\op_{1}$  $\op_{4}$ 
& +   $\p_{5}$  $\op_{2}$  $\op_{3}$ 
& +   $\p_{5}$   $\s_{1}$  $\os_{4}$                           
 \\ 
  +   $\p_{5}$   $\s_5$  $\os_{11}$
& +   $\p_{5}$   $\s_6$  $\os_{12}$
& +   $\p_{5}$   $H_4$  $H_6$
& +   $\p_{5}$   $H_7$  $H_{11}$
& +   $\p_{6}$   $\s_2$  $\os_{3}$                              
\\ 
  +   $\p_{6}$   $\s_7$  $\os_{9}$ 
& +   $\p_{6}$   $\s_8$  $\os_{10}$
& +   $\p_{6}$   $H_3$  $H_6$
& +   $\p_{6}$   $H_8$  $H_{12}$
& +   $\p_{7}$   $\p_{8}$   $\p_{9}$                          
\\ 
  +   $\p_{7}$   $\s_{25}$  $\s_{32}$
& +   $\p_{7}$   $\s_{26}$  $\s_{31}$
& +   $\p_{7}$   $\s_{27}$  $\s_{30}$
& +   $\p_{7}$   $\s_{28}$  $\s_{29}$
& +   $\p_{7}$   $H_{23}$  $H_{26}$                       
 \\
  +   $\p_{7}$   $H_{24}$  $H_{25}$
& +   $\p_{8}$  $\os_{13}$ $\os_{20}$
& +   $\p_{8}$  $\os_{14}$ $\os_{19}$
& +   $\p_{8}$  $\os_{15}$ $\os_{18}$
& +   $\p_{8}$  $\os_{16}$ $\os_{17}$                        
\\
  +   $\p_{8}$   $H_{19}$  $H_{22}$
& +   $\p_{8}$   $H_{20}$  $H_{21}$
& +   $\p_{9}$   $\s_5$   $\s_{12}$
& +   $\p_{9}$   $\s_6$   $\s_{11}$
& +   $\p_{9}$   $\s_7$   $\s_{10}$                                
  \\
  +   $\p_{9}$   $\s_8$   $\s_9$ 
& +   $\p_{9}$   $H_7$  $H_{10}$
& +   $\p_{9}$   $H_8$  $H_9$
& +  $\op_{1}$   $\s_{28}$ $\os_{27}$
& +  $\op_{1}$   $\s_{32}$ $\os_{31}$                          
  \\
  +  $\op_{1}$   $\s_{36}$ $\os_{33}$
& +  $\op_{1}$   $H_1$  $H_3$
& +  $\op_{1}$   $H_{26}$  $H_{30}$
& +  $\op_{2}$   $\s_{27}$ $\os_{28}$
& +  $\op_{2}$   $\s_{31}$ $\os_{32}$                           
 \\
  +  $\op_{2}$   $\s_{35}$ $\os_{34}$
& +  $\op_{2}$   $H_2$  $H_3$
& +  $\op_{2}$   $H_{25}$  $H_{29}$
& +  $\op_{3}$   $\s_{18}$ $\os_{13}$
& +  $\op_{3}$   $\s_{20}$ $\os_{15}$                         
 \\
  +  $\op_{3}$   $\s_{24}$ $\os_{21}$
& +  $\op_{3}$   $H_1$  $H_5$
& +  $\op_{3}$   $H_{18}$  $H_{22}$
& +  $\op_{4}$   $\s_{17}$ $\os_{14}$
& +  $\op_{4}$   $\s_{19}$ $\os_{16}$                        
\\
  +  $\op_{4}$   $\s_{23}$ $\os_{22}$
& +  $\op_{4}$   $H_2$  $H_5$
& +  $\op_{4}$   $H_{17}$  $H_{21}$
& +  $\op_{5}$   $\s_4$  $\os_{1}$ 
& +  $\op_{5}$   $\s_{11}$ $\os_{5}$                     
  \\ 
  +  $\op_{5}$   $\s_{12}$ $\os_{6}$ 
& +  $\op_{5}$   $H_3$  $H_5$
& +  $\op_{5}$   $H_{10}$  $H_{14}$
& +  $\op_{6}$   $\s_3$  $\os_{2}$ 
& +  $\op_{6}$   $\s_9$  $\os_{7}$                          
  \\
  +  $\op_{6}$   $\s_{10}$ $\os_{8}$ 
& +  $\op_{6}$   $H_4$  $H_5$
& +  $\op_{6}$   $H_9$  $H_{13}$
& +  $\op_{7}$  $\op_{8}$  $\op_{9}$ 
& +  $\op_{7}$  $\os_{25}$ $\os_{32}$                 
\\
  +  $\op_{7}$  $\os_{26}$ $\os_{31}$
& +  $\op_{7}$  $\os_{27}$ $\os_{30}$
& +  $\op_{7}$  $\os_{28}$ $\os_{29}$
& +  $\op_{7}$   $H_{27}$  $H_{30}$
& +  $\op_{7}$   $H_{28}$  $H_{29}$                  
  \\
  +  $\op_{8}$   $\s_{13}$  $\s_{20}$
& +  $\op_{8}$   $\s_{14}$  $\s_{19}$
& +  $\op_{8}$   $\s_{15}$  $\s_{18}$
& +  $\op_{8}$   $\s_{16}$  $\s_{17}$
& +  $\op_{8}$   $H_{15}$  $H_{18}$                    
 \\
  +  $\op_{8}$   $H_{16}$  $H_{17}$
& +  $\op_{9}$  $\os_{5}$  $\os_{12}$
& +  $\op_{9}$  $\os_{6}$  $\os_{11}$
& +  $\op_{9}$  $\os_{7}$  $\os_{10}$
& +  $\op_{9}$  $\os_{8}$  $\os_{9}$                       
\\ 
  +  $\op_{9}$   $H_{11}$  $H_{14}$
& +  $\op_{9}$   $H_{12}$  $H_{13}$
& +   $\s_{1}$   $\s_{23}$  $\s_{36}$
& +   $\s_{1}$   $\s_{24}$  $\s_{35}$
& +   $\s_{1}$  $\os_{19}$ $\os_{30}$                         
\\
  +   $\s_{1}$  $\os_{20}$ $\os_{29}$
& +   $\s_{1}$   $H_{21}$  $H_{30}$
& +   $\s_{1}$   $H_{22}$  $H_{29}$
& +   $\s_2$   $\s_{23}$  $\s_{34}$
& +   $\s_2$   $\s_{24}$  $\s_{33}$                                
 \\
  +   $\s_2$  $\os_{17}$ $\os_{32}$
& +   $\s_2$  $\os_{18}$ $\os_{31}$
& +   $\s_2$   $H_{21}$  $H_{28}$
& +   $\s_2$   $H_{22}$  $H_{27}$
& +   $\s_3$   $\s_{21}$  $\s_{36}$                            
\\
  +   $\s_3$   $\s_{22}$  $\s_{35}$
& +   $\s_3$  $\os_{15}$ $\os_{26}$
& +   $\s_3$  $\os_{16}$ $\os_{25}$
& +   $\s_3$   $H_{19}$  $H_{30}$
& +   $\s_3$   $H_{20}$  $H_{29}$
\\
  +   $\s_4$   $\s_{21}$  $\s_{34}$
& +   $\s_4$   $\s_{22}$  $\s_{33}$
& +   $\s_4$  $\os_{13}$ $\os_{28}$
& +   $\s_4$  $\os_{14}$ $\os_{27}$
& +   $\s_4$   $H_{19}$  $H_{28}$
\\
  +   $\s_4$   $H_{20}$  $H_{27}$
& +   $\s_5$   $\s_{17}$  $\s_{33}$
& +   $\s_5$   $\s_{25}$ $\os_{20}$
& +   $\s_5$  $\os_{24}$ $\os_{32}$
& +   $\s_5$  $H_2$  $H_{14}$
\\
  +   $\s_5$   $H_{21}$  $H_{25}$
& +   $\s_6$   $\s_{18}$  $\s_{34}$
& +   $\s_6$   $\s_{26}$ $\os_{19}$
& +   $\s_6$  $\os_{23}$ $\os_{31}$
& +   $\s_6$  $H_1$  $H_{14}$
\\
  +   $\s_6$   $H_{22}$  $H_{26}$
& +   $\s_7$   $\s_{19}$  $\s_{35}$
& +   $\s_7$   $\s_{27}$ $\os_{18}$
& +   $\s_7$  $\os_{24}$ $\os_{30}$
& +   $\s_7$   $H_2$  $H_{13}$
\\
  +   $\s_7$   $H_{21}$  $H_{23}$
& +   $\s_8$   $\s_{20}$  $\s_{36}$
& +   $\s_8$   $\s_{28}$ $\os_{17}$
& +   $\s_8$  $\os_{23}$ $\os_{29}$
& +   $\s_8$  $H_1$  $H_{13}$
\\
  +   $\s_8$   $H_{22}$  $H_{24}$
& +   $\s_9$   $\s_{13}$  $\s_{33}$
& +   $\s_9$   $\s_{29}$ $\os_{16}$
& +   $\s_9$  $\os_{22}$ $\os_{28}$
& +   $\s_9$   $H_2$  $H_{12}$
\\
  +   $\s_9$   $H_{19}$  $H_{25}$
& +   $\s_{10}$  $\s_{14}$  $\s_{34}$
& +   $\s_{10}$  $\s_{30}$ $\os_{15}$
& +   $\s_{10}$ $\os_{21}$ $\os_{27}$
& +   $\s_{10}$ $H_1$ $H_{12}$
\\
  +   $\s_{10}$  $H_{20}$  $H_{26}$
& +   $\s_{11}$  $\s_{15}$  $\s_{35}$
& +   $\s_{11}$  $\s_{31}$ $\os_{14}$
& +   $\s_{11}$ $\os_{22}$ $\os_{26}$
& +   $\s_{11}$  $H_2$  $H_{11}$
\\
  +   $\s_{11}$  $H_{19}$  $H_{23}$
& +   $\s_{12}$  $\s_{16}$  $\s_{36}$
& +   $\s_{12}$  $\s_{32}$ $\os_{13}$
& +   $\s_{12}$ $\os_{21}$ $\os_{25}$
& +   $\s_{12}$ $H_1$  $H_{11}$
\\
  +   $\s_{12}$  $H_{20}$  $H_{24}$
& +   $\s_{13}$  $\s_{26}$ $\os_{4}$ 
& +   $\s_{13}$ $\os_{12}$ $\os_{30}$
& +   $\s_{13}$  $H_4$  $H_{22}$
& +   $\s_{13}$  $H_{13}$  $H_{28}$
\\
  +   $\s_{14}$  $\s_{25}$ $\os_{4}$
& +   $\s_{14}$ $\os_{11}$ $\os_{29}$
& +   $\s_{14}$  $H_4$  $H_{21}$
& +   $\s_{14}$  $H_{13}$  $H_{27}$
& +   $\s_{15}$  $\s_{28}$ $\os_{3}$
\\
  +   $\s_{15}$ $\os_{10}$ $\os_{32}$
& +   $\s_{15}$  $H_3$  $H_{22}$
& +   $\s_{15}$  $H_{14}$  $H_{30}$
& +   $\s_{16}$  $\s_{27}$ $\os_{3}$ 
& +   $\s_{16}$ $\os_{9}$  $\os_{31}$
\\
  +   $\s_{16}$  $H_3$  $H_{21}$
& +   $\s_{16}$  $H_{14}$  $H_{29}$
& +   $\s_{17}$  $\s_{30}$ $\os_{2}$ 
& +   $\s_{17}$ $\os_{8}$  $\os_{26}$
& +   $\s_{17}$  $H_4$  $H_{20}$
\\
  +   $\s_{17}$  $H_{11}$  $H_{28}$
& +   $\s_{18}$  $\s_{29}$ $\os_{2}$
& +   $\s_{18}$ $\os_{7}$  $\os_{25}$
& +   $\s_{18}$  $H_4$  $H_{19}$
& +   $\s_{18}$  $H_{11}$  $H_{27}$
\\
  +   $\s_{19}$  $\s_{32}$ $\os_{1}$
& +   $\s_{19}$ $\os_{6}$  $\os_{28}$
& +   $\s_{19}$  $H_3$  $H_{20}$
& +   $\s_{19}$  $H_{12}$  $H_{30}$
& +   $\s_{20}$  $\s_{31}$ $\os_{1}$
\\ 
  +   $\s_{20}$ $\os_{5}$  $\os_{27}$
& +   $\s_{20}$  $H_3$  $H_{19}$
& +   $\s_{20}$  $H_{12}$  $H_{29}$
& +   $\s_{21}$  $\s_{25}$ $\os_{10}$
& +   $\s_{21}$  $\s_{27}$ $\os_{12}$
\\
  +   $\s_{21}$  $H_{13}$  $H_{26}$
& +   $\s_{21}$  $H_{14}$  $H_{24}$
& +   $\s_{22}$  $\s_{26}$ $\os_{9}$ 
& +   $\s_{22}$  $\s_{28}$ $\os_{11}$
& +   $\s_{22}$  $H_{13}$  $H_{25}$
\\
  +   $\s_{22}$  $H_{14}$  $H_{23}$
& +   $\s_{23}$  $\s_{29}$ $\os_{6}$ 
& +   $\s_{23}$  $\s_{31}$ $\os_{8}$ 
& +   $\s_{23}$  $H_{11}$  $H_{26}$
& +   $\s_{23}$  $H_{12}$  $H_{24}$
\\
  +   $\s_{24}$  $\s_{30}$ $\os_{5}$ 
& +   $\s_{24}$  $\s_{32}$ $\os_{7}$ 
& +   $\s_{24}$  $H_{11}$  $H_{25}$
& +   $\s_{24}$  $H_{12}$  $H_{23}$
& +   $\s_{25}$  $H_6$  $H_{30}$
\\
  +   $\s_{25}$  $H_8$  $H_{20}$
& +   $\s_{26}$  $H_6$  $H_{29}$
& +   $\s_{26}$  $H_8$  $H_{19}$
& +   $\s_{27}$  $H_6$  $H_{28}$
& +   $\s_{27}$  $H_7$  $H_{20}$
\\
  +   $\s_{28}$  $H_6$  $H_{27}$
& +   $\s_{28}$  $H_7$  $H_{19}$
& +   $\s_{29}$  $H_5$  $H_{30}$
& +   $\s_{29}$  $H_{10}$  $H_{22}$
& +   $\s_{30}$  $H_5$  $H_{29}$
\\
  +   $\s_{30}$  $H_{10}$  $H_{21}$
& +   $\s_{31}$  $H_5$  $H_{28}$
& +   $\s_{31}$  $H_9$  $H_{22}$
& +   $\s_{32}$  $H_5$  $H_{27}$
& +   $\s_{32}$  $H_9$  $H_{21}$
\\
  +   $\s_{33}$  $H_8$  $H_{18}$
& +   $\s_{33}$  $H_{10}$  $H_{16}$
& +   $\s_{34}$  $H_8$  $H_{17}$
& +   $\s_{34}$  $H_{10}$  $H_{15}$
& +   $\s_{35}$  $H_7$  $H_{18}$
\\
  +   $\s_{35}$  $H_9$  $H_{16}$
& +   $\s_{36}$  $H_7$  $H_{17}$
& +   $\s_{36}$  $H_9$  $H_{15}$
& +  $\os_{1}$  $\os_{23}$ $\os_{36}$
& +  $\os_{1}$  $\os_{24}$ $\os_{35}$
\\
  +  $\os_{1}$   $H_{15}$  $H_{24}$
& +  $\os_{1}$   $H_{16}$  $H_{23}$
& +  $\os_{2}$  $\os_{23}$ $\os_{34}$
& +  $\os_{2}$  $\os_{24}$ $\os_{33}$
& +  $\os_{2}$   $H_{15}$  $H_{26}$
\\
  +  $\os_{2}$   $H_{16}$  $H_{25}$
& +  $\os_{3}$  $\os_{21}$ $\os_{36}$
& +  $\os_{3}$  $\os_{22}$ $\os_{35}$
& +  $\os_{3}$   $H_{17}$  $H_{24}$
& +  $\os_{3}$   $H_{18}$  $H_{23}$
\\
  +  $\os_{4}$  $\os_{21}$ $\os_{34}$
& +  $\os_{4}$  $\os_{22}$ $\os_{33}$
& +  $\os_{4}$   $H_{17}$  $H_{26}$
& +  $\os_{4}$   $H_{18}$  $H_{25}$
& +  $\os_{5}$  $\os_{17}$ $\os_{33}$
\\
  +  $\os_{5}$  $H_1$  $H_7$
& +  $\os_{5}$   $H_{16}$  $H_{28}$
& +  $\os_{6}$  $\os_{18}$ $\os_{34}$
& +  $\os_{6}$   $H_2$  $H_7$
& +  $\os_{6}$   $H_{15}$  $H_{27}$
\\
  +  $\os_{7}$  $\os_{19}$ $\os_{35}$
& +  $\os_{7}$  $H_1$  $H_8$
& +  $\os_{7}$   $H_{16}$  $H_{30}$
& +  $\os_{8}$  $\os_{20}$ $\os_{36}$
& +  $\os_{8}$   $H_2$  $H_8$
\\
  +  $\os_{8}$   $H_{15}$  $H_{29}$
& +  $\os_{9}$  $\os_{13}$ $\os_{33}$
& +  $\os_{9}$  $H_1$  $H_9$
& +  $\os_{9}$   $H_{18}$  $H_{28}$
& +  $\os_{10}$ $\os_{14}$ $\os_{34}$
\\
  +  $\os_{10}$  $H_2$  $H_9$
& +  $\os_{10}$  $H_{17}$  $H_{27}$
& +  $\os_{11}$ $\os_{15}$ $\os_{35}$
& +  $\os_{11}$ $H_1$  $H_{10}$
& +  $\os_{11}$  $H_{18}$  $H_{30}$
\\
  +  $\os_{12}$ $\os_{16}$ $\os_{36}$
& +  $\os_{12}$  $H_2$  $H_{10}$
& +  $\os_{12}$  $H_{17}$  $H_{29}$
& +  $\os_{13}$  $H_3$  $H_{15}$
& +  $\os_{13}$  $H_8$  $H_{25}$
\\
  +  $\os_{14}$  $H_3$  $H_{16}$
& +  $\os_{14}$  $H_8$  $H_{26}$
& +  $\os_{15}$  $H_4$  $H_{15}$
& +  $\os_{15}$  $H_7$  $H_{23}$
& +  $\os_{16}$  $H_4$  $H_{16}$
\\
  +  $\os_{16}$  $H_7$  $H_{24}$
& +  $\os_{17}$  $H_3$  $H_{17}$
& +  $\os_{17}$  $H_{10}$  $H_{25}$
& +  $\os_{18}$  $H_3$  $H_{18}$
& +  $\os_{18}$  $H_{10}$  $H_{26}$
\\
  +  $\os_{19}$  $H_4$  $H_{17}$
& +  $\os_{19}$  $H_9$  $H_{23}$
& +  $\os_{20}$  $H_4$  $H_{18}$
& +  $\os_{20}$  $H_9$  $H_{24}$
& +  $\os_{21}$  $H_7$  $H_{29}$
\\
  +  $\os_{21}$  $H_8$  $H_{27}$
& +  $\os_{22}$  $H_7$  $H_{30}$
& +  $\os_{22}$  $H_8$  $H_{28}$
& +  $\os_{23}$  $H_9$  $H_{29}$
& +  $\os_{23}$  $H_{10}$  $H_{27}$
\\
  +  $\os_{24}$  $H_9$  $H_{30}$
& +  $\os_{24}$  $H_{10}$  $H_{28}$
& +  $\os_{25}$  $H_5$  $H_{25}$
& +  $\os_{25}$  $H_{14}$  $H_{17}$
& +  $\os_{26}$  $H_5$  $H_{26}$
\\
  +  $\os_{26}$  $H_{14}$  $H_{18}$
& +  $\os_{27}$  $H_5$  $H_{23}$
& +  $\os_{27}$  $H_{13}$  $H_{17}$
& +  $\os_{28}$  $H_5$  $H_{24}$
& +  $\os_{28}$  $H_{13}$  $H_{18}$
\\
  +  $\os_{29}$  $H_6$  $H_{25}$
& +  $\os_{29}$  $H_{12}$  $H_{15}$
& +  $\os_{30}$  $H_6$  $H_{26}$
& +  $\os_{30}$  $H_{12}$  $H_{16}$
& +  $\os_{31}$  $H_6$  $H_{23}$
\\
  +  $\os_{31}$  $H_{11}$  $H_{15}$
& +  $\os_{32}$  $H_6$  $H_{24}$
& +  $\os_{32}$  $H_{11}$  $H_{16}$
& +  $\os_{33}$  $H_{11}$  $H_{21}$
& +  $\os_{33}$  $H_{13}$  $H_{19}$
\\
  +  $\os_{34}$  $H_{11}$  $H_{22}$
& +  $\os_{34}$  $H_{13}$  $H_{20}$
& +  $\os_{35}$  $H_{12}$  $H_{21}$
& +  $\os_{35}$  $H_{14}$  $H_{19}$
& +  $\os_{36}$  $H_{12}$  $H_{22}$
\\
  +  $\os_{36}$  $H_{14}$  $H_{20}$
  &
  &
  &
  &
  \\
\end{supertabular}

\section{Mirror Matter Model Inputs}\label{sec:Mirror_Inputs}
Presented here are the inputs which produced the models in section \ref{sec:disc} that were not included in the discussion of those models.
\begin{flushleft}
\textbf{Table B.a - Mirrored Matter Representations Model 2 - Basis Vectors}
\vspace{3 mm}\\
\begin{tabular}{||c|c||c|c||c|c|c||c|c|c||}
\hline
\hline
\textbf{Sec}&\textbf{N}&$\psi^1$&$\psi^{1*}$&$x^1$
&$y^1$
&$w^1$
&$x^2$
&$y^2$
&$w^2$
\\
\hline
1&2&1&1&1&1&1&1&1&1\\
\hline
S&2&1&1&1&0&0&1&0&0\\
\hline
$b_1$&2&1&1&1&0&0&1&0&0\\
\hline
$b_2$&2&1&1&0&1&0&0&1&0\\
\hline
$b_3$&2&1&1&0&1&0&0&1&0\\
\hline
$\alpha_{1}$&2&1&1&1&0&0&1&0&0\\
\hline
\hline
\end{tabular}

\vspace{3 mm}
\begin{tabular}{||c|c||c|c|c||c|c|c||c|c|c||c|c|c||}
\hline
\hline
\textbf{Sec}&\textbf{N}&$x^3$&$y^3$&$w^3$&$x^4$&$y^4$&$w^4$&$x^5$&$y^5$&$w^5$&$x^6$&$y^6$&$w^6$\\
\hline
1&2&1&1&1&1&1&1&1&1&1&1&1&1\\
\hline
S&2&1&0&0&1&0&0&1&0&0&1&0&0\\
\hline
$b_1$&2&0&1&0&0&1&0&0&1&0&0&1&0\\
\hline
$b_2$&2&1&0&0&1&0&0&0&1&0&0&1&0\\
\hline
$b_3$&2&0&1&0&0&1&0&1&0&0&1&0&0\\
\hline
$\alpha_{1}$&2&1&1&1&1&1&1&1&0&0&1&1&1\\
\hline
\hline
\end{tabular}

\vspace{3 mm}
\begin{tabular}{||c|c||c|c|c|c|c|c|c|c|c|c||c|c|c|c|c|c||}
\hline
\hline
\textbf{Sec}&\textbf{N}&$\overline{\psi}^{1}$&$\overline{\psi}^{1*}$&$\overline{\psi}^2$&$\overline{\psi}^{2*}$&$\overline{\psi}^{3}$&$\overline{\psi}^{3*}$&$\overline{\psi}^{4}$&$\overline{\psi}^{4*}$&$\overline{\psi}^{5}$&$\overline{\psi}^{5*}$&$\overline{\eta}^{1}$&$\overline{\eta}^{1*}$&$\overline{\eta}^{2}$&$\overline{\eta}^{2*}$&$\overline{\eta}^{3}$&$\overline{\eta}^{3*}$\\
\hline
1&2&1&1&1&1&1&1&1&1&1&1&1&1&1&1&1&1\\
\hline
S&2&0&0&0&0&0&0&0&0&0&0&0&0&0&0&0&0\\
\hline
$b_1$&2&1&1&1&1&1&1&1&1&1&1&1&1&0&0&0&0\\
\hline
$b_2$&2&1&1&1&1&1&1&1&1&1&1&0&0&1&1&0&0\\
\hline
$b_3$&2&1&1&1&1&1&1&1&1&1&1&0&0&0&0&1&1\\
\hline
$\alpha_{1}$&2&1&1&1&1&1&1&1&1&1&1&1&1&1&1&1&1\\
\hline
\end{tabular}

\vspace{3 mm}
\begin{tabular}{||c|c||c|c|c|c|c|c||c|c|c|c|c|c||}
\hline
\hline
\textbf{Sec}&\textbf{N}&$\overline{y}^{1}$&$\overline{y}^{2}$&$\overline{y}^{3}$&$\overline{y}^{4}$&$\overline{y}^{5}$&$\overline{y}^{6}$&$\overline{w}^{1}$&$\overline{w}^{2}$&$\overline{w}^{3}$&$\overline{w}^{4}$&$\overline{w}^{5}$&$\overline{w}^{6}$\\
\hline
1&2&1&1&1&1&1&1&1&1&1&1&1&1\\
\hline
S&2&0&0&0&0&0&0&0&0&0&0&0&0\\
\hline
$b_1$&2&0&0&1&1&1&1&0&0&0&0&0&0\\
\hline
$b_2$&2&1&1&0&0&1&1&0&0&0&0&0&0\\
\hline
$b_3$&2&1&1&1&1&0&0&0&0&0&0&0&0\\
\hline
$\alpha_{1}$&2&0&0&1&1&0&1&0&1&1&1&1&1\\
\hline
\end{tabular}

\vspace{3 mm}
\begin{tabular}{||c|c||c|c|c|c|c|c|c|c|c|c|c|c|c|c|c|c||}
\hline
\hline
\textbf{Sec}&\textbf{N}&$\overline{\phi}^{1}$&$\overline{\phi}^{1*}$&$\overline{\phi}^{2}$&$\overline{\phi}^{2*}$&$\overline{\phi}^{3}$&$\overline{\phi}^{3*}$&$\overline{\phi}^{4}$&$\overline{\phi}^{4*}$&$\overline{\phi}^{5}$&$\overline{\phi}^{5*}$&$\overline{\phi}^{6}$&$\overline{\phi}^{6*}$&$\overline{\phi}^{7}$&$\overline{\phi}^{7*}$&$\overline{\phi}^{8}$&$\overline{\phi}^{8*}$\\
\hline
1&2&1&1&1&1&1&1&1&1&1&1&1&1&1&1&1&1\\
\hline
S&2&0&0&0&0&0&0&0&0&0&0&0&0&0&0&0&0\\
\hline
$b_1$&2&0&0&0&0&0&0&0&0&0&0&0&0&0&0&0&0\\
\hline
$b_2$&2&0&0&0&0&0&0&0&0&0&0&0&0&0&0&0&0\\
\hline
$b_3$&2&0&0&0&0&0&0&0&0&0&0&0&0&0&0&0&0\\
\hline
$\alpha_{1}$&2&0&0&0&0&0&0&0&0&0&0&1&1&1&1&1&1\\
\hline
\end{tabular}
\vspace{3 mm}

\textbf{Table B.b - Mirror Matter Representations Model 2 - $k_{ij}$ Matrix}
\vspace{3 mm}\\
\textbf{$k_{ij}\ Matrix\ \times{2}:\ $}\\
\vspace{2 mm}
$\left(\begin{tabular}{c|cccccc}
&1&S&$b_1$&$b_2$&$b_3$&$\alpha_{1}$\\
\hline
1&0&0&2&2&2&2\\
S&0&0&0&0&0&2\\
$b_1$&2&2&2&2&2&2\\
$b_2$&2&2&2&2&2&2\\
$b_3$&2&2&2&2&2&2\\
$\alpha_{1}$&2&2&2&2&2&0\\
\end{tabular}\right)$
\vspace{6 mm}\\

\textbf{Table B.c Mirror Matter Representation Model 3 - $k_{ij}$ Matrix}
\vspace{3 mm}\\
\textbf{$k_{ij}\ Matrix\ \times{6}:\ $}\\
\vspace{2 mm}
$\left(\begin{tabular}{c|ccccc}
&1&$b_1$&$b_2$&$b_3$&$\alpha_{0}$\\
\hline
1&0&6&6&6&4\\
$b_1$&6&6&6&6&4\\
$b_2$&6&6&6&6&2\\
$b_3$&6&6&6&6&2\\
$\alpha_{0}$&6&6&6&6&0\\
\end{tabular}\right)$
\vspace{6 mm}
\\
\newpage
\textbf{Table B.d - Mirror Matter Representations Model 4 - Basis Vectors}
\vspace{2 mm}\\
\begin{tabular}{||c|c||c|c||c|c|c||c|c|c||}
\hline
\hline
\textbf{Sec}&\textbf{N}&$\psi^1$&$\psi^{1*}$&$x^1$
&$y^1$
&$w^1$
&$x^2$
&$y^2$
&$w^2$
\\
\hline
1&2&1&1&1&1&1&1&1&1\\
\hline
S&2&1&1&1&0&0&1&0&0\\
\hline
$b_1$&2&1&1&1&0&0&1&0&0\\
\hline
$b_2$&2&1&1&0&1&0&0&1&0\\
\hline
$b_3$&2&1&1&0&1&0&0&1&0\\
\hline
$\alpha_{1}$&3&1&1&1&0&0&1&0&0\\
\hline
\hline
\end{tabular}
\vspace{3 mm}\\
\begin{tabular}{||c|c||c|c|c||c|c|c||c|c|c||c|c|c||}
\hline
\hline
\textbf{Sec}&\textbf{N}&$x^3$&$y^3$&$w^3$&$x^4$&$y^4$&$w^4$&$x^5$&$y^5$&$w^5$&$x^6$&$y^6$&$w^6$\\
\hline
1&2&1&1&1&1&1&1&1&1&1&1&1&1\\
\hline
S&2&1&0&0&1&0&0&1&0&0&1&0&0\\
\hline
$b_1$&2&0&1&0&0&1&0&0&1&0&0&1&0\\
\hline
$b_2$&2&1&0&0&1&0&0&0&1&0&0&1&0\\
\hline
$b_3$&2&0&1&0&0&1&0&1&0&0&1&0&0\\
\hline
$\alpha_{1}$&3&0&1&0&0&1&0&0&1&0&0&1&0\\
\hline
\hline
\end{tabular}
\vspace{3 mm}\\
\begin{tabular}{||c|c||c|c|c|c|c|c|c|c|c|c||c|c|c|c|c|c||}
\hline
\hline
\textbf{Sec}&\textbf{N}&$\overline{\psi}^{1}$&$\overline{\psi}^{1*}$&$\overline{\psi}^{2}$&$\overline{\psi}^{2*}$&$\overline{\psi}^{3}$&$\overline{\psi}^{3*}$&$\overline{\psi}^{4}$&$\overline{\psi}^{4*}$&$\overline{\psi}^{5}$&$\overline{\psi}^{5*}$&$\overline{\eta}^{1}$&$\overline{\eta}^{1*}$&$\overline{\eta}^{2}$&$\overline{\eta}^{2*}$&$\overline{\eta}^{3}$&$\overline{\eta}^{3*}$\\
\hline
1&2&1&1&1&1&1&1&1&1&1&1&1&1&1&1&1&1\\
\hline
S&2&0&0&0&0&0&0&0&0&0&0&0&0&0&0&0&0\\
\hline
$b_1$&2&1&1&1&1&1&1&1&1&1&1&1&1&0&0&0&0\\
\hline
$b_2$&2&1&1&1&1&1&1&1&1&1&1&0&0&1&1&0&0\\
\hline
$b_3$&2&1&1&1&1&1&1&1&1&1&1&0&0&0&0&1&1\\
\hline
$\alpha_{1}$&3&1&1&1&1&1&1&1&1&1&1&0&0&0&0&1&1\\
\hline
\end{tabular}
\vspace{3 mm}\\
\begin{tabular}{||c|c||c|c|c|c|c|c||c|c|c|c|c|c||}
\hline
\hline
\textbf{Sec}&\textbf{N}&$\overline{y}^{1}$&$\overline{y}^{2}$&$\overline{y}^{3}$&$\overline{y}^{4}$&$\overline{y}^{5}$&$\overline{y}^{6}$&$\overline{w}^{1}$&$\overline{w}^{2}$&$\overline{w}^{3}$&$\overline{w}^{4}$&$\overline{w}^{5}$&$\overline{w}^{6}$\\
\hline
1&2&1&1&1&1&1&1&1&1&1&1&1&1\\
\hline
S&2&0&0&0&0&0&0&0&0&0&0&0&0\\
\hline
$b_1$&2&0&0&1&1&1&1&0&0&0&0&0&0\\
\hline
$b_2$&2&1&1&0&0&1&1&0&0&0&0&0&0\\
\hline
$b_3$&2&1&1&1&1&0&0&0&0&0&0&0&0\\
\hline
$\alpha_{1}$&3&0&0&0&0&1&1&0&0&0&0&0&0\\
\hline
\end{tabular}
\vspace{3 mm}\\
\begin{tabular}{||c|c||c|c|c|c|c|c|c|c|c|c|c|c|c|c|c|c||}
\hline
\hline
\textbf{Sec}&\textbf{N}&$\overline{\phi}^{1}$&$\overline{\phi}^{1*}$&$\overline{\phi}^{2}$&$\overline{\phi}^{2*}$&$\overline{\phi}^{3}$&$\overline{\phi}^{3*}$&$\overline{\phi}^{4}$&$\overline{\phi}^{4*}$&$\overline{\phi}^{5}$&$\overline{\phi}^{5*}$&$\overline{\phi}^{6}$&$\overline{\phi}^{6*}$&$\overline{\phi}^{7}$&$\overline{\phi}^{7*}$&$\overline{\phi}^{8}$&$\overline{\phi}^{8*}$\\
\hline
1&2&1&1&1&1&1&1&1&1&1&1&1&1&1&1&1&1\\
\hline
S&2&0&0&0&0&0&0&0&0&0&0&0&0&0&0&0&0\\
\hline
$b_1$&2&0&0&0&0&0&0&0&0&0&0&0&0&0&0&0&0\\
\hline
$b_2$&2&0&0&0&0&0&0&0&0&0&0&0&0&0&0&0&0\\
\hline
$b_3$&2&0&0&0&0&0&0&0&0&0&0&0&0&0&0&0&0\\
\hline
$\alpha_{1}$&3&0&0&0&0&0&0&1&1&1&1&1&1&1&1&1&1\\
\hline
\end{tabular}
\vspace{6 mm}\\

\textbf{Table B.e - Mirror Matter Representation - Model 4 - $k_{ij}$ Matrix}
\textbf{$k_{ij}\ Matrix\ \times{6}:\ $}\\
\vspace{2 mm}
$\left(\begin{tabular}{c|cccccc}
&1&S&$b_1$&$b_2$&$b_3$&$\alpha_{1}$\\
\hline
1&0&0&6&6&6&0\\
S&0&0&0&0&0&0\\
$b_1$&6&6&6&6&6&0\\
$b_2$&6&6&6&6&6&0\\
$b_3$&6&6&6&6&6&6\\
$\alpha_{1}$&0&6&0&6&0&4\\
\end{tabular}\right)$

\end{flushleft}
}%END APPENDIX

\end{document}